\documentclass[12pt]{article}
\usepackage{amssymb}
\usepackage{amsmath}
\usepackage{amsfonts,amsthm}
\usepackage{graphicx}
\usepackage{xcolor} 
\usepackage{natbib}
\usepackage{booktabs} 

\newtheorem{proposition}{Proposition}

\newcommand{\bm}[1]{\mbox{\boldmath $#1$}}
\newcommand{\argmin}{\mathop{\rm argmin}}

\newcommand{\bms}[1]{\mbox{\scriptsize\boldmath $#1$}}

\title{\bf Minimum Copula Divergence \\ for Robust Estimation}
\author{{Shinto Eguchi}  {and} {Shogo Kato } \\
{ Institute of Statistical Mathematics}, Tokyo}
\date{\null}

\begin{document}
\bibliographystyle{jae.bst}

\setcitestyle{authoryear,open={(},close={)}}
\maketitle

\begin{abstract}
This paper introduces a robust estimation framework based solely on the copula function. We begin by introducing a family of divergence measures tailored for copulas, including the \(\alpha\)-, \(\beta\)-, and \(\gamma\)-copula divergences, which quantify the discrepancy between a parametric copula model and an empirical copula derived from data independently of marginal specifications. Using these divergence measures, we propose the minimum copula divergence estimator (MCDE), an estimation method that minimizes the divergence between the model and the empirical copula. The framework proves particularly effective in addressing model misspecifications and analyzing heavy-tailed data, where traditional methods such as the maximum likelihood estimator (MLE) may fail. Theoretical results show that common copula families, including Archimedean and elliptical copulas, satisfy conditions ensuring the boundedness of divergence-based estimators, thereby guaranteeing the robustness of MCDE, especially under extreme observations. Numerical examples further underscore MCDE's ability to adapt to varying dependence structures.
\end{abstract}

\newpage

\section{Introduction}

Copulas are mathematical functions that capture the dependence structure between random variables, allowing for the construction of multivariate distributions by combining univariate marginal distributions with a copula that describes their interdependencies. This separation is particularly useful in modeling complex relationships where variables exhibit nonlinear or asymmetric dependencies. In finance, copulas are extensively used to model and manage risks associated with multiple financial assets. They enable the assessment of joint default probabilities in credit risk management, the pricing of complex derivatives like collateralized debt obligations, and the evaluation of portfolio risk by capturing tail dependencies that traditional correlation measures might miss. For instance, copulas have been applied to monitor market risk in basket products and to measure credit risk in large pools of loans  {(see, e.g., \cite{mcneil2015})}. In engineering, copulas facilitate the analysis of systems where component dependencies significantly impact overall reliability. They have been applied in structural engineering to assess the reliability of tower and tower-line systems under varying loads, such as wind or earthquakes. In environmental sciences, copulas are employed to model and analyze dependencies between various environmental factors. For example, they have been used to study the joint distribution of temperature and precipitation in the Mediterranean region, providing insights into climate patterns. See, for example,  \cite{bevilacqua2024} for an application to spatial copula modeling.

Despite these successful applications, several challenges remain. Traditional estimation methods such as the maximum likelihood estimator often prove highly sensitive to model misspecifications and extreme observations, particularly in high-dimensional settings or when data exhibit heavy tails.  {It is important to clarify the notion of robustness discussed in this paper. Since our estimation procedure is based on the empirical copula, which depends only on coordinate-wise ranks, it is inherently robust against outliers or misspecification in the one-dimensional marginal distributions. Our focus, therefore, is on robustness against misspecification of the dependence structure itself, such as when the true copula is a contaminated version of the parametric model. Under such misspecification, the estimated parameter $\bm\theta$ is interpreted as the parameter of the model copula that is ``closest'' to the true underlying copula, as measured by the chosen divergence.} These issues underscore the need for robust methodologies that can isolate and accurately capture the underlying dependence structure even under adverse conditions. Our proposed framework, based on copula divergence will directly address these challenges by bypassing marginal assumptions and mitigating the influence of outliers, thereby enhancing the reliability of dependence modeling across diverse fields. For this, we introduce a family of divergence functionals defined on the space of copula functions, called copula divergence. Specifically, as variants of copula divergence,  the $\alpha$-copula,  $\beta$-copula, and $\gamma$-copula divergences are presented,  where the parameters $\alpha$,  $\beta$  and $\gamma$ represent the respective power exponents characterizing each divergence. For a parametric copula model,  the estimator is proposed by minimization for the copula divergence between the model copula and the empirical copula with respect to the parameter, which is called the minimum copula divergence estimator (MCDE).  {This approach falls within the semiparametric estimation framework for copula models, where marginal distributions are treated nonparametrically, an idea extensively explored by \cite{Genest1995} and \cite{tsukahara2005}, among others.} In this way, MCDE is a rank statistic as a functional of the empirical copula function like Kendall's $\tau$ and  Spearman's $\rho$, and it differs fundamentally from the MLE. MCDE can exhibit greater robustness to extreme observations; while the log-likelihood (even the pseudo-likelihood) can be more sensitive if the copula density blows up near corners or edges.  {It is well-known that the pseudo-likelihood estimator performs very well in many settings when the model is correctly specified, and sieve estimators can achieve semiparametric efficiency \citep{chen2006efficient, hofert2019elements}.}
However, it can suffer from large bias or variance in smaller samples, especially under heavy tails. MCDE might lose a bit of asymptotic efficiency  relative to MLE in the perfect-model scenario, but can be more robust under misspecification or contamination.  {Our contribution is not to propose the semiparametric approach itself, but to introduce a unified and flexible class of divergence-based estimators (MCDEs) within this framework and to analyze their robustness properties through a novel power-boundedness condition.}
In computational aspects, MLE, for many copula families, requires evaluating and differentiating the copula density. Some integrals or boundary terms can complicate the process, especially in higher dimensions. MCDE typically needs only to evaluate values of the copula model function, and an associated divergence functional. For many Archimedean or elliptical copulas, the model function is simpler to handle with no partial derivatives  than the corresponding density function. The rest of this paper is organized as follows. Section 2 discusses various power divergences. Section 3 formalizes the minimum copula divergence estimator and its properties. Section 4 explores influence functions and robustness. Section 5 provides examples of Archimedean copula.


\section{Copula models}

Let us consider a statistical inference about  copula functions. Copulas are functions that describe the dependence structure between random variables. They separate the joint distribution of random variables into two parts:
 the marginal distributions of each random variable
and the copula function capturing dependence. Sklar's theorem states that for a vector of random variables 
\(\bm{X} = (X_1, X_2, \ldots, X_d)\) with joint distribution \(F\) 
and marginals \(F_1, F_2, \ldots, F_d\), there exists a copula \(C\) such that
\[
    F(\bm x) 
    = C\bigl(F_1(x_1), F_2(x_2), \ldots, F_d(x_d)\bigr).
\]
In particular, for continuous marginals, \(C\) is unique.  See \cite{nelsen2006} for comprehensive introduction. We  consider a copula model 
\[
{\cal M}=\{C_{\bms\theta}(\bm u):\bm \theta\in{\Theta}\},
\]
where $\bm\theta$ is an unknown parameter of a parameter space $\Theta$ that is an open subset of $\mathbb R^m$. See \cite{joe2014} for detailed treatment of copulas to model complex dependence structures.
\subsection{Common copula models}

We highlight two important families of copula models: 
 Archimedean copulas and elliptical copulas. \vspace{2mm}

\noindent{\bf   Archimedean  Copulas}:
They form a widely used class characterized 
by a \emph{generator} function \(\psi\), see \cite{Genest1993}. They are often appealing 
due to their analytic simplicity and ability to model both tail 
dependence and asymmetry (depending on the generator). An Archimedean copula \(C\) in \(d\) dimensions is defined by
\[
   C(\bm u) 
   = \psi^{[-1]}\bigl(\psi(u_1) + \cdots + \psi(u_d)\bigr),
\]
where 
\(\psi:[0,1] \to [0,\infty]\) is a continuous, strictly decreasing 
\emph{generator function} with \(\psi(1) = 0\), and 
\(\psi^{[-1]}\) is the (generalized) inverse function of \(\psi\).  
Crucially, \(\psi'(u) < 0\) to ensure the required properties.
Many well-known one-parameter Archimedean copulas can be written 
in terms of their generator \(\psi_\theta\). \begin{itemize}
    \item \textbf{Clayton Copula}:
    \[
       \psi_\theta(t) =\frac{1} {\theta}\bigl(t^{-\theta} - 1\bigr), 
       \quad \theta > 0; \quad
       C_{\theta}(u_1,\dots,u_d) 
       = \biggl(
           1 + \sum_{k=1}^d \bigl(u_k^{-\theta} - 1\bigr)
         \biggr)^{-1/\theta}.
\]

    \item \textbf{Gumbel Copula}:
    \[
       \psi_\theta(t) = (-\ln t)^\theta, 
       \quad \theta \ge 1; \quad
       C_{\theta}(u_1,\dots,u_d) 
       = \exp\Bigl\{-\Bigl(
            (-\ln u_1)^\theta + \cdots + (-\ln u_d)^\theta
         \Bigr)^{1/\theta}\Bigr\}.
\]

    \item \textbf{Frank Copula}:
    \[
       \psi_\theta(t) = -\ln\biggl(\frac{e^{-\theta t} - 1}{e^{-\theta} - 1}\biggr),
       \quad \theta \in \mathbb{R}\setminus\{0\};
\]
    \[
       C_{\theta}(u_1,\dots,u_d) 
       = -\frac{1}{\theta}\ln \Bigl[
           1 - \bigl(1 - e^{-\theta}\bigr)
           \prod_{k=1}^d \frac{e^{-\theta u_k}-1}{e^{-\theta}-1}
         \Bigr].
\]

    \item \textbf{Joe Copula}:
    \[
       \psi_\theta(t) = -\log\bigl\{1-(1-t)^\theta\bigr\}, 
       \quad \theta \ge 1; \quad
       C_{\theta}(\bm{u})
       = 1 - \Bigl[(1 - u_1)^\theta + \cdots + (1 - u_d)^\theta 
            - \prod_{i=1}^d (1 - u_i)^\theta\Bigr]^{1/\theta}.
\]
\end{itemize}

\noindent{\bf  Elliptical Copulas}

\vspace{2mm}

Elliptical distributions  induce to  copula functions arisen from such as the 
multivariate normal (Gaussian) or multivariate \(t\)-distribution. They are useful for modeling symmetric dependence structures. 

\begin{itemize}
    \item \textbf{Gaussian Copula}:
    Derived from the multivariate normal distribution, the Gaussian copula 
    is defined using the standard normal cumulative distribution function (CDF) \(\Phi\) and the multivariate normal CDF 
    \(\Phi_\Sigma\) (with mean \(\mathbf{0}\) and correlation matrix \(\Sigma\)):
    \[
       C_{\Sigma}(u_1,\dots,u_d) 
       = \Phi_{\Sigma}\bigl(\Phi^{-1}(u_1),\dots,\Phi^{-1}(u_d)\bigr).
\]
    Here, \(\Phi^{-1}\) is the standard normal quantile function.
    \item \textbf{Student-\(t\) Copula}:
    Arising from the multivariate \(t\)-distribution, the Student-\(t\) copula 
    uses the univariate \(t\)-distribution CDF \(t_\nu\) (with \(\nu\) degrees of freedom) 
    and its multivariate counterpart \(t_{\nu,\Sigma}\):
    \[
       C_{\nu,\Sigma}(u_1,\dots,u_d) 
       = t_{\nu,\Sigma}\bigl(t_{\nu}^{-1}(u_1),\dots,t_{\nu}^{-1}(u_d)\bigr)
    \]
    Its heavier tails facilitate modeling of data that exhibit stronger tail dependence.
\end{itemize}

Building on these examples, the remainder of this work investigates 
divergence-based methods for estimating copula parameters. In particular, we explore how divergences can yield robust and efficient 
procedures for a variety of copula families, including both Archimedean and elliptical. For this we introduce copula divergence, typically the power copula divergence and its variants.
\subsection{Power-bounded copula model}\label{power}

\begin{figure}[htbp]
\centering
  \includegraphics[width=\textwidth]{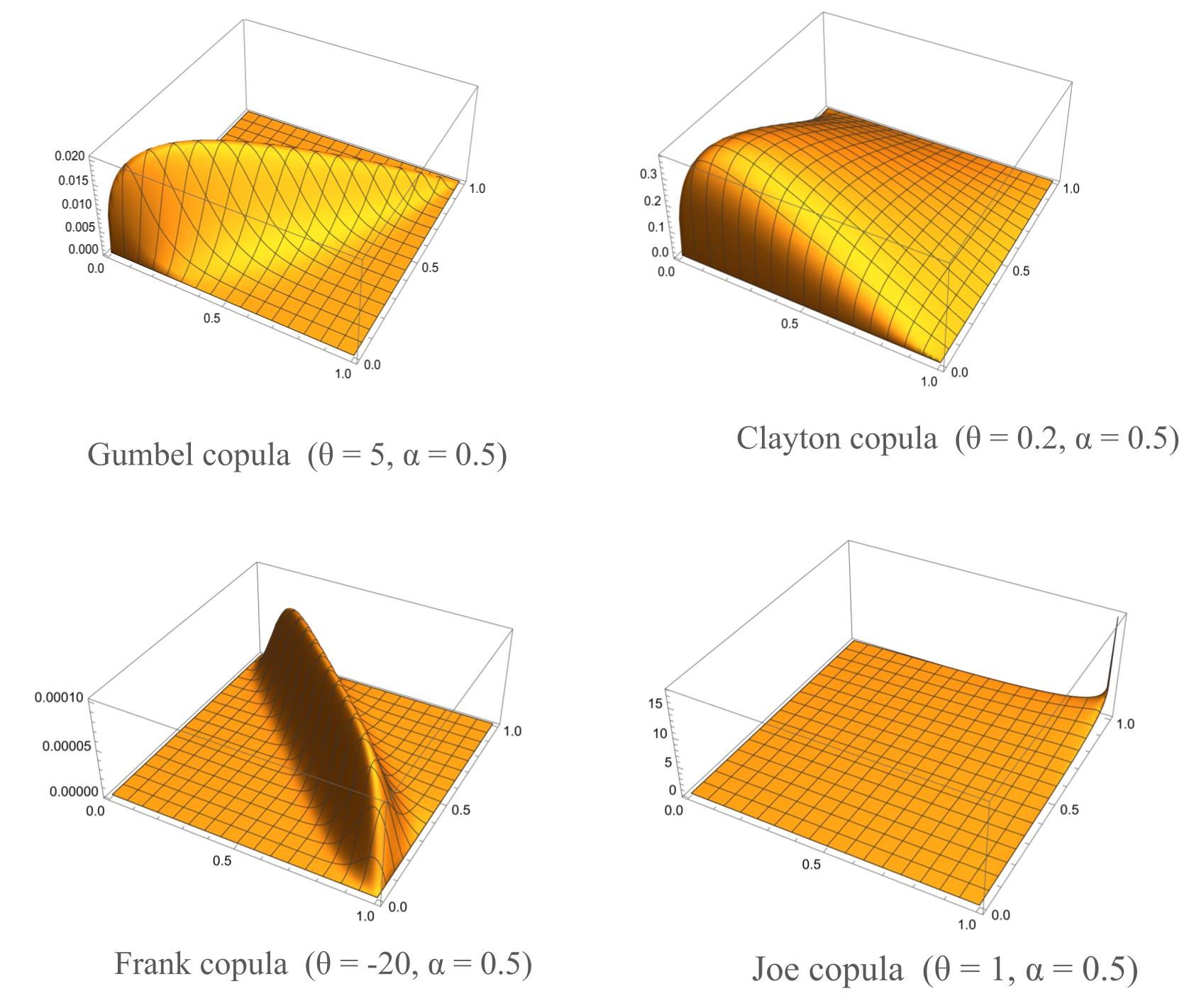}
 \caption{ 3D-plots of the derivative functions.}
\label{copula1}
\end{figure}

We discuss a sensible assumption of lower tail behaviors for a copula model. For simplicity, we suppose a  bivariate copula model $C_{\bms\theta}(u,v)$, and the discussion will be easily extended to multivariate cases. We say a copula model $C_{\bms\theta}(u,v)$ is  \emph{power-bounded} if 
$C_{\bms\theta}(u,v)$ satisfies
\begin{align}\label{result}
\sup_{(u,v)\in[0,1]^2}  \| C_{\bms\theta}(u,v)^\alpha \,\nabla_{\bms\theta}\,\log C_{\bms\theta}(u,v)\|<\infty
\end{align}
for any $\alpha>0$, where $\nabla_{\bms\theta}$ denotes a gradient vector for $\bm\theta$. Usually,   \(\nabla_{\bms\theta}\,\log C_{\bms\theta}(u,v)\) may 
blow up near \((u,v)=(0,0)\)  {noting $\lim_{(u,v)\to(0,0)} C_{\bms\theta}(u,v)= 0$.}
See \cite{charpentier(2009)} for related discussion for  copula behaviors near the corners of the unit square. The condition \eqref{result} indicates that multiplying by \(C_{\bms\theta}(u,v)^\alpha\) effectively controls the potential blow-up of the log-derivative near the boundaries. In fact, this power-boundedness condition is satisfied for many popular Archimedean copula families. Table \ref{table1} summarizes the supremum values of these powered log-derivatives for several typical copula functions. Although the derivations in Appendix 1 involve intricate calculus, the essential idea is straightforward: one can show that 
 {the limit of $C_{\bms\theta}(u,u)^\alpha \nabla_{\bms\theta}\,\log C_{\bms\theta}(u,u)$ as $u \to 0$ is zero}
for any \(\alpha > 0\). This bounded behavior is also clearly illustrated in Figure \ref{copula1}, which displays the log-derivative function for the specific case \(\alpha=\frac{1}{2}\).
\begin{table}[htbp]
\centering
\caption{ {Supremum of $\| C_{\bms\theta}^\alpha \nabla_{\bms\theta} \log C_{\bms\theta} \|$ near the origin.}}
\label{table1}
\vspace{3mm}
\begin{tabular}{|l|c|c|c|}
\hline
     & $\alpha>0$ & $\alpha=0$ & $\alpha<0$ \\
\hline
     Clayton   & $0$        & $\infty$   & $\infty$   \\
     Gumbel    & $0$        & $\infty$   & $\infty$   \\
     Frank     & $0$        & \;$\frac{1}{\theta}\!-\!\frac{1}{\exp(\theta)-1}$ & $\infty$   \\
     Joe       & $0$        & $\frac{1}{\theta}$   & $\infty$ \\  
\hline                              
\end{tabular}
\end{table}

We next consider a Gaussian copula model that is defined as:
$$C_\theta(u,v) = \Phi_\theta(\Phi^{-1}(u), \Phi^{-1}(v)),  $$
where $\Phi_\theta$ is a bivariate standard normal cumulative distribution function with correlation parameter $\bm\theta$, and $\Phi^{-1}$ is the inverse of the univariate standard normal cumulative distribution function. Let $C_\theta(u,v) = \Phi_\theta\!\bigl(\Phi^{-1}(u),\,\Phi^{-1}(v)\bigr)$ 
be the bivariate Gaussian copula with correlation parameter $\theta$, 
where $\Phi_\theta$ is the bivariate standard normal CDF and $\Phi$ 
is the univariate standard normal CDF. Then, as $(u,v)\to (0,0)$, 
the following limits hold:
\[
\bigl|\nabla_\theta \log C_\theta(u,v)\bigr|
\,\to\,
\infty
\quad\text{and}\quad
\bigl|C_\theta(u,v)^\alpha 
\,\nabla_\theta \log C_\theta(u,v)\bigr|
\,\to\, 0
\quad\text{for all }\alpha>0.
\]
The proof is given in Appendix 2.
We remark the diverging property for the log-derivative of the Gaussian copula comes from the exponential decay of the Gaussian distribution in the left tail. On the other hand, the Student t copula function $C_\theta(u,v)=t_{\nu,\theta}(t_\nu(u),t_\nu(v))$ has only a polynomial decay.   Hence, $|\nabla_\theta \log C_\theta(u,v)|$ becomes bounded.
\section{Copula divergence}

We introduce a family of copula-based divergences.
This differs from standard density-based divergences like the Kullback-Leibler divergence
and other density divergences. See \cite{{basu1998robust},{jones2001},{Fujisawa2008},{eguchi2022minimum}} for many variants of density divergence.
In principle, copula divergences focus {exclusively} on the dependence structure among variables, while density divergences measure discrepancies in the entire joint distribution--including both dependence and marginals. Copula divergences remove the influence of marginal distributions and \emph{directly} quantify how the shape of the dependence structures differ. This can be more robust when the margins are unknown or contaminated, and it directly addresses questions such as ``How does the association (e.g., tail dependence) differ between two copulas?''

Let $\cal C$ be the set of all the copula functions defined on a hypercube ${\cal H}=[0,1]^d$. A  functional $D$ defined on ${\cal C}\times{\cal C}$ is called a divergence  if $D$ satisfies
\begin{align}\nonumber
D(C_0,C_1)\geq0
\end{align}
with equality if and only if $C_0(\bm u)=C_1(\bm u)$ for almost surely $\bm u$ of $\cal H$. Let us discuss three types of power divergence as follows.
\begin{itemize}
\item {\bf  $\alpha$-copula divergence}:
\begin{align}\nonumber
D_{\alpha}(C_0,C_1)=\frac{1}{\alpha(1-\alpha)} \int_{\cal H} {C_0}(\bm u)
\bigg\{1-\Big(\frac{{C_1}(\bm u)}{{C_0}(\bm u)}\Big)^\alpha+\alpha
\Big(\frac{{C_1}(\bm u)}{{C_0}(\bm u)}-1\Big)\bigg\}{\rm d}C_0(\bm u).
\end{align}
\item {\bf $\beta$-copula divergence}:
\begin{align}\nonumber
D_{\beta}(C_0,C_1)=\int_{\cal H}\Big\{
\frac{{C_0}(\bm u)^{\beta+1}}{\beta(\beta+1)}+\frac{{C_1}(\bm u)^{\beta+1}}{\beta+1}-\frac{{{C_0}(\bm u)}{{C_1}(\bm u)}^\beta}{\beta} \Big\} {\rm d}C_0(\bm u).
\end{align}

\item{\bf $\gamma$-copula divergence}:
\begin{align}\nonumber
D_\gamma(C_0,C_1)=-\frac{1}{\gamma}
\frac{  \int_{\cal H}{{C_0}(\bm u)}{{C_1}(\bm u)}^\gamma {\rm d}C_0(\bm u)}
{\big( \int_{\cal H} {C_1}(\bm u)^{\gamma+1}{\rm d}C_0(\bm u)\big)^{ \frac{\gamma}{\gamma+1}}}+\frac{1}{\gamma}
{\Bigl( \int_{\cal H} {C_1}(\bm u)^{\gamma+1}{\rm d}C_0(\bm u)\Bigr)^{ \frac{1}{\gamma+1}}}.
\end{align}
\end{itemize}
Here `${\rm d}C_{0}(\bm u)$' in the above equations represents  integration with respect to the measure induced by $C_{0}(\bm u)$. For the $\alpha$-copula divergence, if $\alpha=\frac{1}{2}$, then it is reduced to the Hellinger squared distance
\begin{align}\label{Hellinger}
D_{\rm H}(C_0,C_1)=  \int_{\cal H} 
\Big\{{C_1}(\bm u)^{\frac{1}{2}}-{C_0}(\bm u)^{\frac{1}{2}}\Big\}^2{\rm d}C_0(\bm u).
\end{align}
If $\alpha\to 0$, then it is reduced to 
the extended KL-divergence
\begin{align}\nonumber
D_{\rm eKL}(C_0,C_1)= \int_{\cal H}\Big\{{{C_0}(\bm u)}\log \frac{{C_0}(\bm u)}{{C_1}(\bm u)}
+{C}_1(\bm u)-{C}_0(\bm u)\Big\}{\rm d}C_0(\bm u).
\end{align}
Let $\psi$ be a strictly convex function defined on $\mathbb R$ with $\psi^\prime(1)=0$ and $\psi^{\prime\prime}(1)=1$. Then, $\psi$-divergence $D_\psi(C_0,C_1)$ is defined by
\begin{align}\nonumber
\int \psi\bigg(\frac{C_1(\bm u)}{C_0(\bm u)}\bigg){C_0(\bm u)}{\rm d} C_0(\bm u).
\end{align}
If we choose as $$\psi(t)=\frac{1-t^\alpha}{\alpha(1-\alpha)}-\frac{1-t}{1-\alpha},$$ then the $\psi$-divergence is nothing but the $\alpha$-copula divergence. See Csiszar (1967) for the case of density divergence. 

For the $\beta$-copula divergence, if $\beta=1$, it is reduced to  {a half of the Cram\'{e}r-von-Mises  divergence, an estimator studied in \cite{tsukahara2005}}. \begin{align}\label{CM}
D_{\rm CM}(C_0,C_1)=  \int
\big\{{{C_0}(\bm u)}-{C_{1}(\bm u)}\big\}^2d C_0(\bm u).
\end{align}
If $\beta\to 0$, it is reduced to
the extended KL-divergence $D_{\rm eKL}(C_0,C_1)$. If $\beta\to -1$, it is reduced to
the Itakura-Saito divergence
\begin{align}\nonumber
D_{\rm IS}(C_0,C_1)= \int_{\cal H}\Big\{\log \frac{{C_0}(\bm u)}{{C_1}(\bm u)}
+\frac{{C}_1(\bm u)}{{C}_0(\bm u)}-1\Big\}{\rm d}C_0(\bm u).
\end{align}
Moreover, the $\beta$-copula divergence has a natural extension from the power function to a convex function as a generator function. Let $U$ be a strictly increasing and convex function defined on $\mathbb R$. Then, $U$-divergence $D_U(C_0,C_1)$ is defined by
\begin{align}\nonumber
\int \big[U(\xi(C_1(\bm u)))-U(\xi(C_0(\bm u))) -C_0(\bm u)\{\xi(C_1(\bm u))-\xi(C_0(\bm u))\}\big]{\rm d}C_0(\bm u),
\end{align}
where $\xi$ is the inverse of the derivative function of $U$. It is noted that $D_U(C_0,C_1)\geq0$ with equality if and only if $C_0=C_1$ since
$U(B)-U(A)\geq u(A)(B-A)$ from the assumption of $U$. If $U(t)=(1+\beta t)^{\frac{\beta+1}{\beta}}/(\beta+1)$, then 
$\xi(s)=(s^\beta-1)/\beta$, so that the $U$-divergence is reduced to the $\beta$-power divergence.
For the $\beta$-copula divergence, if $\gamma=1$, 
\begin{align}\nonumber
D_\gamma (C_0,C_1)\big|_{\gamma=1} =-
\frac{ \int_{\cal H}{{C_0}(\bm u)}{{C_1}(\bm u)} {\rm d}C_0(\bm u)- \sqrt{\int_{\cal H}{{C_0}(\bm u)}^2 {\rm d}C_0((\bm u)}\sqrt{\int_{\cal H}{{C_1}(\bm u)} {\rm d}C_0(\bm u)}}
{\sqrt{ \int_{\cal H} {C_1}(\bm u)^{2}{\rm d}C_0(\bm u)}}.
\end{align}
This is associated with the Cauchy-Schwartz inequality.
See \cite{{Fujisawa2008},{eguchi2011projective},{eguchi2024}} for the $\gamma$-density divergence and the geometric perspectives.

We remark that, if $\gamma$ is taken a limit to $-1$, then the $\gamma$-copula diagonal entropy is written as
\begin{align}\nonumber
\lim_{\gamma\rightarrow-1}H_\gamma (C_0,C_0) &= \lim_{\gamma\rightarrow-1}
-\frac{1}{\gamma}\left(\int_{\cal H}{{C_0}(\bm u)}^{\gamma+1} {\rm d}C_0(\bm u)\right)^\frac{1}{\gamma+1}
\\[3mm]\nonumber
&= \exp\left(\int_{\cal H}\log{{C_0}(\bm u)}{\rm d}C_0(\bm u)\right).
\end{align}
This is nothing but the geometric mean of $C_0(\bm u)$, say $H_{\rm GM}(C_0)$. Thus, 
\begin{align}\nonumber
\lim_{\gamma\rightarrow-1} D_\gamma (C_0,C_1)&=
{  \int_{\cal H}\frac{{C_0}(\bm u)}{{C_1}(\bm u)}{\rm d}C_0(\bm u)}
H_{\rm GM}(C_1)-H_{\rm GM}(C_0),
\end{align}
which is called the geometric-mean divergence. For $\gamma=-2$, then the $\gamma$-copula diagonal entropy is a half of the harmonic mean of $C_0(\bm u)$:
\begin{align}\nonumber
H_\gamma (C_0,C_0)\big|_{\gamma=-2}&=\frac{1}{2}
\left(  \int_{\cal H}{C_0}(\bm u)^{-1}{\rm d}C_0(\bm u)\right)^{-1}, \mbox{ say } H_{\rm HM}(C_0)
\end{align}
and hence the harmonic mean divergence is given by
\begin{align}\nonumber
D_{\rm HM} (C_0,C_1)&=\frac{1}{2}
{  \int_{\cal H}{{C_0}(\bm u)}{{C_1}(\bm u)}^{-2} {\rm d}C_0(\bm u)}
{H_{\rm HM}(C_1)^2}-{H_{\rm HM}(C_0)}
\end{align}

In general, copula divergences focus only on differences in dependence structures. Their derivation is often numerically simpler for many copula models, since they require evaluating  the copula function rather than integrating the copula density function. Naturally, we expect that  copula divergences are robust to marginal misspecification, since they measure structure on \([0,1]^d\)  {pseudo-observations}. In this work, we explore the estimation methods based on copula divergences.
\subsection{Minimum copula divergence}

We apply copula divergences to estimate a parameter in a copula model ${\cal M}=\{C_{\bms\theta}(\bm u):\bm \theta\in{\Theta}\}$. Copula divergences are specifically designed to measure how two {dependence} structures differ, irrespective of marginal changes discussed in the preceding section. Hence, the advantageous property can be transferred to such copula-based estimators.
Let us give a mathematical formulation for proposed estimators. For a random sample $(\bm X_1,...,\bm X_n)$, the  {pseudo-observations} $(\bm U_1,...,\bm U_n)$ are defined by  {\citep{ghoudi2004empirical}}
\begin{align}\label{pseudo}
    \bm U_i =\frac{1}{n+1} \Big(\sum_{j=1}^n \mathbb I(X_{j1}\leq X_{i1}),\cdots, \sum_{j=1}^n \mathbb I(X_{jd}\leq X_{id})\Big)
\end{align}
and the empirical copula function is defined by
\begin{align}\label{emp-copula}
\hat{C}(\bm u)=\frac{1}{n+1}\sum_{i=1}^n \prod_{j=1}^d {\mathbb I}( U_{ij}\leq u_j),
\end{align}
where $\mathbb I$ represents the indicator function. Let $D(C_0,C_1)$ be a copula divergence.
Thus,  the minimum copula divergence estimator (MCDE) for ${\bm\theta}$ is defined by
\begin{align}\label{mcde}
          \hat{\bm \theta}_D=\argmin_{\bms\theta\in\Theta} D(\hat{C},C_{\bms\theta}).
\end{align}

The empirical copula converges weakly to a mean-zero Gaussian process:
\[
  \sqrt{n}\,\bigl(\widehat{C}(\bm{u}) - C(\bm{u})\bigr)
    \Longrightarrow   
  W(\bm{u}),
  \quad \bm{u}\in [0,1]^d,
\]
where \(W\) is  a general Gaussian process with the mean \(\bf0\) and covariance given by  
\[
  \mathbb{E}\bigl[\,W(\bm{u})\,W(\bm{v})\bigr]
   = 
  \Gamma\bigl(\bm{u},\,\bm{v}\bigr)
   = 
  C\!\bigl(\bm{u} \wedge \bm{v}\bigr)
   - 
  C(\bm{u})\,C(\bm{v}),
\]
where \(\bm{u} \wedge \bm{v}\) denotes the coordinatewise minimum \(\bigl(\min(u_1,v_1),\dots,\min(u_d,v_d)\bigr)\). See Van der Vaart and Wellner (1996) or Segers (2012) for rigorous discussion and wide perspectives. Hence the limit is a kind of  multivariate Brownian bridge governed by the true copula \(C\), one might call it a {copula-based Brownian sheet}. We define a functional
\[
   \Lambda(C)
   \;=\;
   \argmin_{{\bms\theta}\in \Theta}\,D\bigl(C, C_{{\bms\theta}}\bigr).
\]
Under regularity conditions (identifiability, interior solution, differentiability of \(D\), etc.), it can 
be applied to the functional delta method for  $\Lambda(C)$ at $C=\hat C$ noting $\hat {\bm\theta}_D=\Lambda(\hat C)$.  {Assuming the data-generating process follows the specified model, i.e., $C = C_{\bms\theta^*}$ for some true value $\bm\theta^* \in \Theta$,} by the continuous mapping theorem, see, e.g., {Billingsley (1999)},  \(   \widehat{{\bm\theta}}_D   \;\Longrightarrow\;   {\bm\theta}^*\)
since  \(\widehat{C}\to C_{{\bms\theta}^*}\) in probability in \(\ell^\infty([0,1]^d)\). Moreover, $\sqrt{n}(\hat {\bm\theta}_D-{\bm\theta})$ weakly converges to a normal distribution with mean $\bf0$ and covariance $\Sigma_D$. The covariance matrix $\Sigma_D$ is derived by the sandwich formula as a typical example for the M-estimator. The concrete form will be given for specific examples of copula divergence $D$. We will use a learning algorithm  to find the numerical minimizer for 
a MCDE. Routinely, a stochastic gradient algorithm is suggested as
\begin{align}\nonumber
{\bm\theta} \leftarrow {\bm\theta} -\alpha \nabla_{\bms\theta}
D(\hat C, C_{\bms\theta}).
\end{align}
 {where $\alpha$ is the learning rate.}
The computational complexity is usually reasonable rather than that for the MLE accompanying with the evaluation of the copula density functions as discussed later.  {One advantage of MCDE is that, when defining divergences purely in terms of copulas (rather than their densities), one avoids complicated integrals. The construction of the empirical copula \(\hat{C}\) bypasses the need to integrate or differentiate a parametric copula density. However, this advantage depends on the tractability of the copula CDF itself. For copula families with analytically simple CDFs, like many Archimedean copulas, this approach offers numerical stability and computational efficiency. We acknowledge that for other families where the density is computationally simpler than the CDF (e.g., vine copulas or high-dimensional elliptical copulas), density-based methods may be more practical.}
It is worthwhile to emphasize how the  {pseudo-observations} come from the empirical distribution of each margin, making MCDE a purely rank-based approach. Thus, MCDE has an advantage such that any scaling or monotonic transformations of the original data do not affect the estimated copula parameter. The present proposal is closely related to the copula divergence discussed in 
\cite{de-keyser2024}, which relies on the copula density functions. It can be referred to as the kernel-based divergence to estimate copulas, see \cite{alquier2023} for statistical performance. By working with \(\hat{C}\) directly, the estimation has a semiparametric flavor: the marginals are replaced by their empirical distributions \(F_j(x_j)\), and only the dependence structure \(C_{\bms\theta}\) is modeled parametrically. We focus on the loss function defined by a copula divergence $D(C_0,C_1)$, i.e., \( L_D(\bm\theta)=D(\hat C, C_{\bms\theta})\). Typically, there are  loss functions by three type of copula divergences as follows:

\begin{itemize}
\item{ \bf  {$\alpha$}-power loss}:   
 \begin{align}\nonumber
L_{\alpha}(\bm\theta)=
\frac{1}{\alpha(1-\alpha)} \sum_{i=1}^n 
\big\{-{{C_{\bms\theta}}(\bm U_i)^\alpha}{{\hat C}(\bm U_i)}^{1-\alpha}+\alpha
{{C_{\bms\theta}}(\bm U_i)}\big\},
\end{align}
and  $\hat{\bm\theta}_\alpha=\argmin_{\bms\theta\in\Theta}L_{\alpha}(\bm\theta)$ is called $\alpha$-MCDE, where $\{\bm U_i\}$ are observed  {pseudo-observations} defined in \eqref{pseudo}. The {$\alpha$}-estimating equation is given by 
$\bm S_\alpha({\bm\theta})= \sum_{i=1}^n \bm S_\alpha(\bm U_i,\bm\theta)={\bf0}$, where
\begin{align}\label{alpha-est}
\bm S_\alpha(\bm u,{\bm\theta})=\frac{1}{1-\alpha}  
{{C_{\bms\theta}}(\bm u)^\alpha}\big\{{{C_{\bms\theta}}(\bm u)}^{1-\alpha}-{{\hat C}(\bm u)^{1-\alpha}}\big\}\nabla_{\bms\theta} \log C_{\bms\theta}(\bm u).
\end{align}

\item{\bf {$\beta$}-power loss}: 
\begin{align}\nonumber
L_{\beta}(\bm\theta)=
-\frac{1}{\beta}  \sum_{i=1}^n{{\hat C}(\bm U_i)}{{C_{\bms\theta}}(\bm U_i)}^\beta
+\frac{1}{\beta+1} \sum_{i=1} {C_{\bms\theta}}(\bm U_i)^{\beta+1},
\end{align}
and  $\hat{\bm\theta}_\beta=\argmin_{\bms\theta\in\Theta}L_{\beta}(\bm\theta)$ is called $\beta$-MCDE. The {$\beta$}-estimating equation is given by 
$\bm S_\beta({\bm\theta})= \sum_{i=1}^n \bm S_\beta(\bm U_i,\bm\theta)={\bf0}$, where
\begin{align}\label{beta-est}
\bm S_\beta(\bm u,{\bm\theta})= 
 {C_{\bms\theta}(\bm u)}^\beta\big\{{C_{\bms\theta}(\bm u)} -{\hat{C}(\bm u)}\big\}\nabla_{\bms\theta} \log C_{\bms\theta}(\bm u).
\end{align}

\item{ \bf  {$\gamma$}-power loss}: 
\begin{align}\nonumber
L_\gamma({\bm\theta})= -\frac{1}{\gamma}\frac{ \sum_{i=1}^n {\hat{C}(\bm U_i)}{C_{\bms\theta}(\bm U_i)}^\gamma}
{\ \ \big( \sum_{i=1}^n {C}_{\bms\theta}(\bm U_i)^{\gamma+1}\big)^{\frac{\gamma}{\gamma+1}}},
\end{align}
and  $\hat{\bm\theta}_\gamma=\argmin_{\bms\theta\in\Theta}L_{\gamma}(\bm\theta)$ is called $\gamma$-MCDE. The estimating equation is given by
$\bm S_\gamma({\bm\theta})= \sum_{i=1}^n \bm S_\gamma(\bm U_i,\bm\theta)={\bf0}$, where
\begin{align}\label{gamma-est}
\bm S_\gamma(\bm u,{\bm\theta}) =
 {C_{\bms\theta}(\bm u)}^\gamma\big\{\hat w_{\gamma,{\bms\theta}}\; {C_{\bms\theta}(\bm u)} -{\hat{C}(\bm u)}\big\}\nabla_{\bms\theta} \log C_{\bms\theta}(\bm u).
\end{align}
Here
\begin{align}\nonumber
\hat w_{\gamma,{\bms\theta}}=\frac{ \sum_{i=1}^n{\hat{C}(\bm U_i)}{C_{\bms\theta}(\bm U_i)}^\gamma}
{\ \   \sum_{i=1}^n {C}_{\bms\theta}(\bm U_i)^{\gamma+1}}.
\end{align}

\end{itemize}
The $\alpha$-estimating function \eqref{alpha-est} consists of  three multiplicative factors: \( C_{\bms{\theta}}(\bm{U}_i)^\alpha\) 
gives a {weight} to  each observation \(\bm{U}_i\) according to the model's own predicted copula value; \( \{C_{\bms{\theta}}(\bm{U}_i)^{1-\alpha} -  \hat{C}(\bm{U}_i)^{1-\alpha}\}\) is  the difference between
the model value and the empirical value in the \((1-\alpha)\) power sense; $\nabla_{\bms\theta} \log C_{\bms\theta}(\bm U_i)$ gives the direction or the gradient for the estimating function. When \(\alpha\) is small, there is less penalty imposed in regions where \(C_{\bms{\theta}}(\bm{U}_i)\) is very small, which can yield a degree of robustness near the boundary \((\bm{u}\approx \bm{0})\). When \(\alpha\) is larger, the function places more importance on getting the model's values correct even in small-copula regions, potentially improving efficiency under ideal conditions but also becoming more sensitive to boundary outliers. Similarly, the $F$-loss function is associated with the $F$-divergence:
 \begin{align}\nonumber
L_{F}(\bm\theta)=
 \sum_{i=1}^n 
  F\bigg(\frac{C_{\bms\theta}(\bm U_i)}{\hat C(\bm U_i)}\bigg){\hat C(\bm U_i)} 
\end{align}
up to the constant. The estimating equation is given by
\begin{align}\nonumber
\bm S_F({\bm\theta})= \sum_{i=1}^n 
  F^{'}\bigg(\frac{C_{\bms\theta}(\bm U_i)}{\hat C(\bm U_i)}\bigg)
 \nabla_{\bms\theta}{C_{\bms\theta}(\bm U_i)}.
\end{align}

On the other hand, the {$\beta$}-estimating equation $\bm S_\beta({\bm\theta})$ is a linear functional of the empirical copula function $\hat C$ and is decomposed into three factors similar to $\alpha$-estimating equation. In particular, $\beta$ can be negative to give {more} weight to points where 
the copula value is small and {less} weight otherwise. This can sometimes be beneficial if one wants to place emphasis on tail regions. Similarly, the $U$-loss function is given by
 \begin{align}\nonumber
L_{U}(\bm\theta)= \sum_{i=1}^n 
\left[ U\big(\xi({C_{\bms\theta}(\bm U_i)})\big)
{\hat C(\bm U_i)}\xi\big({C_{\bms\theta}(\bm U_i)}\big)\right],
\end{align}
and the estimating function is given by
\begin{align}\nonumber
\bm S_U({\bm\theta})= 
 \sum_{i=1}^n 
\left[{C_{\bms\theta}(\bm U_i)}-{\hat C(\bm U_i)}\right]
\nabla_{\bms\theta} \xi\big({C_{\bms\theta}(\bm U_i)}\big).
\end{align}

The $\gamma$-estimating function \eqref{gamma-est} is the same as $\beta$-copula estimating function \eqref{beta-est} except for the extra weight $\hat w_{\gamma,{\bms\theta}}$ if the power exponents $\gamma$ and $\beta$ match. The extra weight is the ratio of the cross and diagonal entropy, $H_\gamma(\hat C,C_{\bms\theta})/H_\gamma( C_{\bms\theta},C_{\bms\theta})$, in which it provides an additional \emph{global} scale adaptation
noting $\hat w_{\gamma,{\bms\theta}}\approx 1$ when $\hat C=C_{\bms\theta}$. See \cite{Fujisawa2008} for detailed discussion of the super robustness as density divergence.
We explore the asymptotic distributions of three types of power MCDEs under
a correctly specified model.

\begin{proposition}\label{prop1}
 {Assume that the copula model $C_{\bms\theta}(\bm u)$ is twice continuously differentiable with respect to $\bm\theta$ and satisfies other standard regularity conditions for M-estimators.}
Assume that $\hat C(\bm u)$ is the empirical copula  defined in \eqref{emp-copula} generated from a copula model $C_{\bms\theta}(\bm u)$. Let $\hat{\bm\theta}_\alpha$, $\hat{\bm\theta}_\beta$, and $\hat{\bm\theta}_\gamma$ be the $\alpha$-MCDE,
the $\beta$-MCDE, and the $\gamma$-MCDE, respectively.
Then,
\begin{itemize}
\item[]$\rm (i)$. $\ \sqrt{n}(\hat{\bm\theta}_\alpha-\bm\theta)\;\Longrightarrow\; N\bigl({\bf0},\Sigma_0(\bm\theta)\bigr)$,

\item[]$\rm (ii)$. $\ 
\sqrt{n}(\hat{\bm\theta}_\beta-\bm\theta)\;\Longrightarrow\; N\bigl({\bf0},\Sigma_\beta(\bm\theta)\bigr)$,

\item[]$\rm (iii)$. 
$\ \sqrt{n}(\hat{\bm\theta}_\gamma-\bm\theta)\;\Longrightarrow\; N\bigl({\bf0},\Sigma_\gamma(\bm\theta)\bigr)$.

\end{itemize}
Here, $\Sigma_x(\bm\theta)=A_x({\bm\theta})^{-1}B_x({\bm\theta})A_x({\bm\theta})^{-1}$
for $x=0, \beta, \gamma$,  where            
\begin{align}\label{A-theta}
 A_x(\bm\theta)&=\int C_{\bms\theta}(\bm u)^{x+1} \nabla_{\bms\theta} \log C_{\bms\theta}(\bm u) \otimes \nabla_{\bms\theta} \log C_{\bms\theta}(\bm u) {\rm d}C_{\bms\theta}(\bm u);
\end{align}
\begin{align}\label{B-theta}
\begin{split}
B_x({\bm\theta})&=\int\!\!\int C_{\bms\theta}(\bm u)^{2x}(C_{\bms\theta}(\bm u \wedge \bm v)-C_{\bms\theta}(\bm u)C_{\bms\theta}(\bm v)) \\
&\qquad \nabla_{\bms\theta} \log C_{\bms\theta}(\bm u) \otimes \nabla_{\bms\theta} \log C_{\bms\theta}(\bm v) {\rm d}C_{\bms\theta}(\bm u){\rm d}C_{\bms\theta}(\bm v).
\end{split}
\end{align}     

\end{proposition}
Proof is given in Appendix 3, which is based on the standard discussion for deriving the asymptotics of M-estimators such as the sandwich formula and Slutsky's theorem. We note the asymptotic distribution for the $\alpha$-MCDE does not depend on $\alpha$. This is closely related with such a property that the minimum $\alpha$-density divergent estimator 
has a unique asymptotic distribution independent of $\alpha$, see \cite{eguchi1983}. As for the $\beta$-MCDE, when $\beta=0$, the asymptotic variance equals that of the $\alpha$-MCDE. When the power exponents $\beta$ and $\gamma$ equal, then the asymptotic variance for the $\gamma$-MCDE equals that of the $\beta$-MCDE. This is because the extra weight $\hat w_{\gamma,{\bms\theta}}$ converges to $1$ in probability.
\subsection{Robust minimum copula divergence}

We discuss  how the class of MCDEs can be framed and compared with both the MLE and minimum \emph{density} divergence estimators. We focus on the methodological motivations, theoretical implications, and practical ways to strengthen the argument in favor of MCDE. The copula-based approach, justified by Sklar's theorem, naturally separates a joint distribution into marginal distributions and a copula capturing dependence. The MCDE is appealing because it focuses solely on the copula component-marginal distributions (and their densities) need not be explicitly specified or estimated. In contrast, density-based approaches such as the MLE or minimum density divergence estimators often involve specifying or approximating the full joint density \( f_{\bms\theta}(x) \). For clarifying this, we give a foundational assumption for the copula model, which could correspond to that for the density model such as exponential family. The key idea is to allow sufficiently flexible behaviors for $\nabla_{\bms\theta} \log C_{\bms\theta}(\bm u)$ of the copula functions in the tail.
As introduced in the preceding section, 
we have discussed a few forms of power copula divergences
with power exponents \(\alpha\), \(\beta\), and \(\gamma\)
and the asymptotic distributions under a correctly specified model. Each choice yields a different copula-based loss function and estimating equation. This unifies common metrics (e.g., Hellinger, Kullback--Leibler, Itakura--Saito) under a single umbrella and provides a range of options for balancing robustness and efficiency. An important aspect of choosing a suitable divergence is controlling the estimator's sensitivity to outliers, particularly in the tails.
Let us investigate statistical properties for MCDEs:
All three estimating functions for $\alpha$-, $\beta$-, and $\gamma$-MCDEs have the following properties:
Therefore, these inequalities are immediate because the contrast term is bounded by $2$.
\begin{proposition}\label{prop2}
Assume  a copula model $C_{\bms\theta}(u,v)$ is  power-bounded as defined in \eqref{result}. Let  $\bm S_\alpha(\bm u,\bm{\theta})$,  $\bm S_\beta(\bm u,\bm{\theta})$,  and $\bm S_\gamma(\bm u,\bm{\theta})$ be estimating functions defined in \eqref{alpha-est}, \eqref{beta-est},
and \eqref{gamma-est}. Then,
\begin{itemize}
\item[]$\rm (i)$. $\bm S_\alpha(\bm u,\bm{\theta})$
is uniformly bounded on $[0,1]^d$ if $0<\alpha<1$.

\item[]$\rm (ii)$. $\bm S_\beta(\bm u,\bm{\theta})$ 
is uniformly bounded on $[0,1]^d$ if $\beta>0$.

\item[]$\rm (iii)$. $\bm S_\gamma(\bm u,\bm{\theta})$
is uniformly bounded on $[0,1]^d$ if $\gamma>0$.
\end{itemize}
\end{proposition}
Proof is given in Appendix 4.
This implies the influence functions of the three MCDEs with $\alpha\in(0,1),\, \beta\in(0,\infty),\, \gamma\in(0,\infty)$
are all  bounded, and hence their MCDEs are led to be qualitatively robust. To compare the usual estimators including the maximum likelihood estimator
we will have numerical experiments in several settings of simulated data generations. The behaviors of  MCDEs should be investigated  in the presence of misspecification, in particular,  outliers  in the upper or lower tail.
 {
\subsection{Data-driven selection of the divergence parameter}
By selecting different \(\alpha\), \(\beta\), or \(\gamma\) parameters, one can tune how the estimator reacts to misspecification or extreme observations. Here is a natural question about which value of the power exponent is optimal? For this, we discuss a data-driven selection of the exponent by cross-validation for a practical situation, where it is not known whether the underlying distribution follows a correctly specified model, or a misspecified model.
Consider a selection for the optimal value of $\beta$ to get good estimator in the family of $\beta$-MCDEs. Let ${\cal D}=\{\bm U_i:1\leq i\leq n\}$, and
 \( \mathcal{D} \) is randomly divided into \( k \) folds:
   \[
   \mathcal{D} = \bigcup_{j=1}^k \mathcal{D}_j \quad \text{with} \quad \mathcal{D}_i \cap \mathcal{D}_j = \emptyset \text{ for } i \neq j.
\]
   Each fold \( \mathcal{D}_j \) contains approximately \({n}/{k} \) samples. We use \( \mathcal{D} \setminus \mathcal{D}_j \)  as the training set and
use \( \mathcal{D}_j \) as the validation set recursively in $j, 1\leq j\leq k$ in the $k$-fold cross validation. Let $ \hat{\bm\theta}_{(-j)}({\beta}) = \argmin_{\bms\theta\in\Theta}L_{(-j)}({\beta},\bm\theta)
$, where $\beta$ is in a grid set of candidates, say $\Omega=\{\beta_1,...,\beta_m \}$. Here 
\begin{align}\nonumber
L_{(-j)}({\beta},\bm\theta)=
  -\sum_{\bms u\in{\mathcal{D} \setminus \mathcal{D}_j}}\left[
\frac{{\hat C}_{(-j)}(\bm u){{C_{\bms\theta}}(\bm u)}^\beta}{\beta}
-\frac{{C_{\bms\theta}}(\bm u)^{\beta+1}}{\beta+1}\right], 
\end{align}
where ${\hat C}_{(-j)}(\bm u)$ is the empirical copula based on ${\mathcal{D} \setminus \mathcal{D}_j}$. Finally, for the empirical copula $\hat C^{(j)}$ is based on ${\cal D}_j$ we evaluate $C_{\hat{\bms\theta}_{(-j)}({\beta})}$ as a predictor for $\hat C^{(j)}$
by a fixed $\beta$-copula divergence with $\beta=0.1$, say  $D_{0}(C_0,C_1)$, as an out-of-sample anchor loss. In this way, we define the optimal $\beta$ as
\begin{align}\nonumber
\hat{\beta}^{\rm opt}= \argmin_{\beta\in \Omega} 
\sum_{j=1}^kD_{0}(\hat C^{(j)},C_{\hat{\bms\theta}_{(-j)}({\beta})})
\end{align}
It is worthwhile to note that $\hat C^{(j)}$ and $C_{\hat{\bms\theta}_{(-j)}({\beta})}$
are statistically independent by construction of ${\cal D}_j$'s. The optimal $\beta$ can escape from over-learning  thanks to the partition of ${\cal D}_j$ and
${\cal D} \setminus{\cal D}_j$. A simple numerical experiment will be given in a subsequent discussion.
}

\subsection{ {Robustness properties and comparison with other methods}}

We discuss the characteristics for MCDEs to compare the usual estimators under a copula model. Typically, the MLE is defined on the model of probability density functions induced from the copula model:  The  density function of $\bm X$  is given by 
\[
f_{\bms\theta}(\bm x)=c_{\bms\theta}(F_1(x_1),...,F_d(x_d))f_1(x_1)\cdots f_d(x_d), 
\]
where $F_j(x_j)$'s  are the cumulative distribution functions (CDFs) with the probability density functions $f_j(x_j)$'s, and
\begin{align}\label{pdf}
c_{\bms\theta}(\bm u)=\frac{\partial^d}{\partial u_1 \cdots \partial u_d}C_{\bms\theta}(u_1,...,u_d).
\end{align}
In this way, the negative log-likelihood function can be written as
$L_{\rm ML}(\bm\theta)=-\sum_{i=1}^n \log f_{\bms\theta}(\bm X_i)$ if 
the CDF $F_j$'s  are known. In effect, there are a few variants of MLEs depending on the way to estimate for 
marginal CDFs. The approach to imposes parametric forms on both margins and the copula is
guaranteed to be more efficient if everything is correctly specified. However,  it is more susceptible to misspecification errors in either the margins or copula leading to bias, especially in heavy-tailed scenarios. We focus on the semiparametric MLE defined by minimization of 
\(
L_{\rm SML}(\bm\theta)=-\sum_{i=1}^n \log c_{\bms\theta}(\bm U_i)
\)
where $\bm U_i$'s are  {pseudo-observations} defined in \eqref{pseudo}. This approach ignores direct parametric assumptions on the margins. 
Instead, it uses empirical distribution functions \(\hat F_j(x_j)\) to transform \(\bm X_i\) to \(\bm U_i\). It is more flexible than fully parametric MLE because we do not impose parametric forms on each marginal distribution. The estimation function is given by
\begin{align}\nonumber
\sum_{i=1}^n \nabla_{\bms\theta} \log c_{\bms\theta}(\bm U_i).
\end{align}
The semiparametric MLEs are widely used for estimating the parameter $\bm\theta$ for the asymptotic efficiency with a technically complicated correction, however, it has a weak point for the robustness. This is because the function $\nabla_{\bms\theta} \log c_\theta(\bm u)$ is often unbounded in $\bm u$ in common copula models. For example, consider the Clayton copula model, in which the density function is given by
 \[
   c_\theta(u,v)
  =
   (\theta + 1)(uv)^{-\theta - 1}\bigl(u^{-\theta} + v^{-\theta} - 1\bigr)^{-2 - \tfrac{1}{\theta}}.
\]
This implies
\[
 \nabla_\theta \log  c_\theta(u,v)
  =
 -\log (u)-\log(v)+\frac{\log(u^{-\theta} + v^{-\theta} - 1)}{\theta^2} 
+\Big(2+\frac{1}{\theta}\Big)
\frac{u^{-\theta}\log(u) + v^{-\theta}\log(v)}{u^{-\theta} + v^{-\theta} - 1}
 \]
up to a constant. Hence,  we observe $|\nabla_\theta \log  c_\theta(u,v)|\rightarrow\infty$ as $(u,v)\rightarrow (0,0)$
since the log-derivative  can involve terms such as $\log(u)$, $\log(v)$, $u^{-\theta}$, $v^{-\theta}$ and so on. Alternatively, density-based  divergences can capture full differences  between two distributions encompassing both marginals and dependence. Even if one needs to focus only on dependence, then a density divergence can be a comprehensive choice--but often at the expense of more complexity and potentially less robustness if the marginals are uncertain. For example, the $\beta$-copula loss functions for a parametric copula density function $c_{\bms\theta}(u)$ can be considered as
\begin{align}\label{loss1}
-\frac{1}{\beta}\frac{1}{n} \sum_{i=1}^n
{c_{\bms\theta}(\bm U_i)}^\beta+\frac{1}{\beta+1}\int_{\cal H}{c}_{\bms\theta}(\bm u)^{\beta+1}d\bm u.
\end{align}
It can provide a robust estimation, however, the multiple integral in the second term has no closed form even for common copula models. For this, there needs numerical integration procedures like MCMC sampling each step in the learning algorithm to find the minimum of the loss function in $\bm\theta$.  {Furthermore, the condition for the robustness of this density-based estimator is more complicated than that for the MCDE discussed here.}
In this way, we will explore further aspects for MCDE rather than the minimum density divergence estimator. With MCDE, the parametric copula \emph{distribution} function \(C_{{\bms\theta}}\) is required rather than its density \(c_{{\bms\theta}}\). Numerically, this is often much simpler since we only need to evaluate \(C_{{\bms\theta}}(\bm{U}_i)\) at  {pseudo-observations}, without having to differentiate or integrate the copula function over \([0,1]^d\). The MLE for copulas often comes from assuming both marginal distributions and a parametric form for the dependence structure, or from a pseudo-likelihood approach if marginals are nonparametric. This can be quite sensitive to tail misspecification.
Under correct specification, the MLE may be asymptotically efficient. However, it can fail badly under heavy tails or outlier contamination. Divergence-based estimators (particularly those that are robust) may exhibit smaller bias or variance under model misspecification, at the cost of a small efficiency loss in the ideal case.
\section{Numerical experiments}

\begin{figure}[htbp]
\centering
\includegraphics[width=\textwidth]{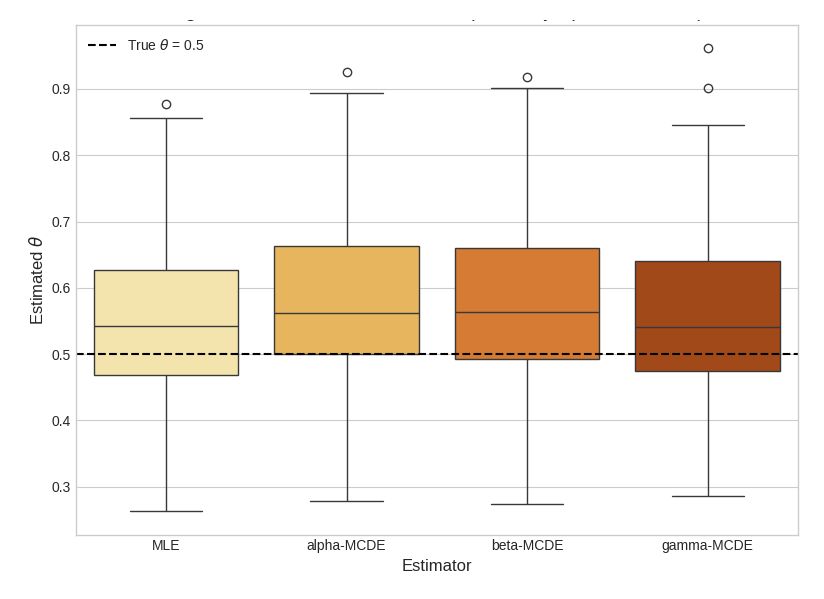}
\caption{Distribution of estimated $\theta$ across 50 repetitions under the correctly specified Clayton copula model}
\label{SMI11}
\end{figure}

Let us have numerical experiments of small scale.
Consider a Clayton copula model: 
\begin{align}\label{Clayto2}
  C_{\theta}(u,v)=\bigl(u^{-{\theta}}+v^{-{\theta}}-1\bigr)^{-\frac{1}{\theta}}
\end{align}
for $\theta>0$.  {First, data is generated from a correctly specified of $C_\theta(u,v)$ with $200$ observations,
where $\theta$ is set as $0.5$. Thus, we find five estimators:  the semiparametric MLE, $\alpha$-MCDE ($\alpha=0.1$), $\beta$-MCDE ($\beta=0.1$) and $\gamma$-MCDE ($\gamma=0.1$). The result of the numerical evaluation for these estimators with repetitions $50$ is given 
in Table \ref{SIM1} and Figure \ref{SMI11}.}
The semiparametric MLE has the best performance among four estimators in the sense of the root mean squared error (RMSE), in which the three MCDEs have almost equivalent performance. The tendency can be observed for other choice for these power exponents $\alpha$, $\beta$, and $\gamma$.

\begin{table}[htbp]
\centering
\caption{Correctly specified case}\label{SIM1}
\vspace{3mm}
\begin{tabular}{|l|cc|c|}
\hline
     &Mean & RMSE  \\
\hline
    MLE   &  {0.5508} &  {0.1361} \\
    $\alpha$-MCDE    &  {0.5770} &  {0.1503} \\
 $\beta$-MCDE  &  {0.5749} &  {0.1498} \\
$\gamma$-MCDE  &  {0.5590} &  {0.1435} \\
\hline
\end{tabular}
\end{table}

\begin{figure}[h]
\centering
\includegraphics[width=\textwidth]{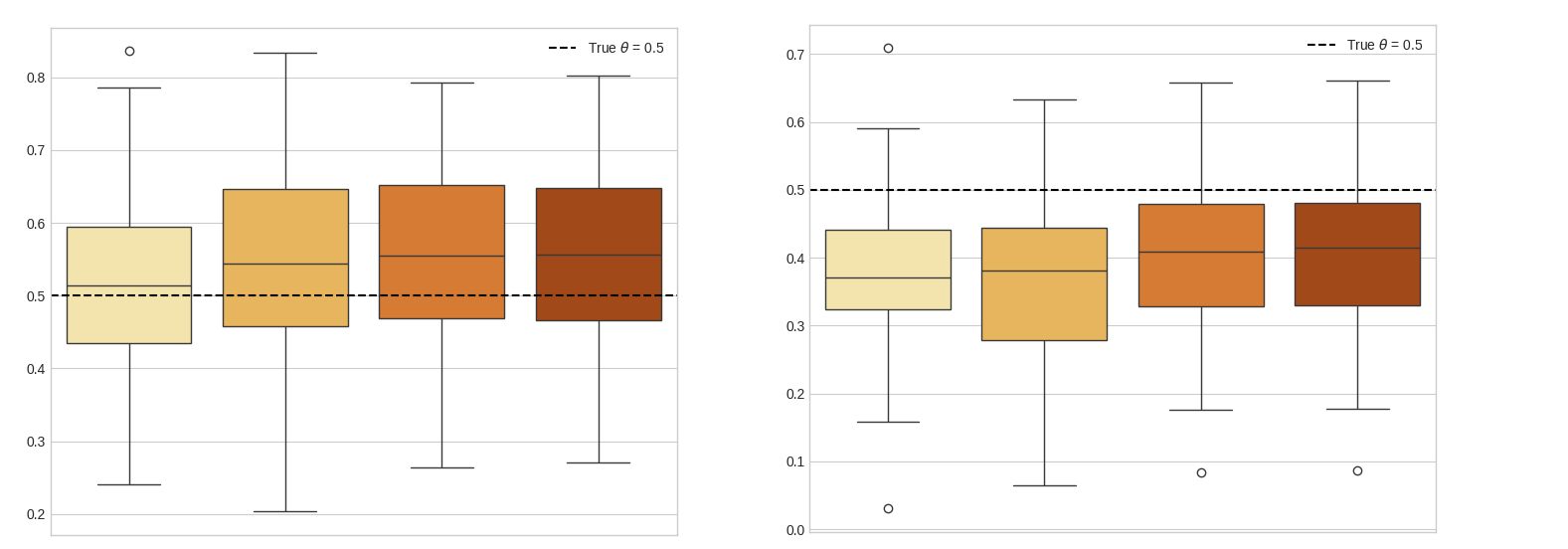}
\caption{ {Plots of estimated $\theta$ under the two misspecified models}}
\label{SMI22}
\end{figure}

\bigskip
Secondly, we now consider a type I of the model misspecified  as
\begin{align}\label{mis}
C_{\rm mis}^{(\rm I)}(u,v)=
(1-\pi)C_{\theta}(u,v)+\pi C_{*}(u,v),
\end{align}
where $\pi$ is set as $0.025$ and $C_*(u,v)$ is a Student t copula function with correlation $-0.5$ and $5$ degrees of freedom. Data is generated from the misspecified model $C_{\rm mis}(u,v)$ with $200$ observations with the true $\theta=0.5$. 
The result of the numerical evaluation for these estimators with $100$ repetitions is given 
in Table \ref{SIM2} and Figure \ref{SMI22}. Under the misspecified model, the semiparametric  MLE still tries to fit a single Clayton copula to data that partly come from a Student t copula. Consequently, its estimate of $\theta$ is pulled away from the nominal truth 
$\theta=0.5$. 
The robust estimators $\alpha$-, $\beta$-, and $\gamma$-MCDEs, fixed as $\alpha=\beta=\gamma=0.1$, downweight (in different ways) outliers or small subsets of data that do not follow the assumed model. As a result, these estimators exhibit:
Smaller bias relative to MLE and significantly smaller RMSE.
 { We now discuss a model where the marginal distributions are contaminated. We generate data $(X_1, X_2)$ where the dependence is governed by the Clayton copula $C_\theta$ with $\theta=0.5$, but the first marginal is a mixture $(1-\pi)N(0,1) + \pi N(5,1)$ with $\pi=0.05$. Pseudo-observations are then generated from this dataset.}
This setting introduces another type of misspecification, causing the MLE to be biased. By contrast, all variants of the MCDE estimators, being rank-based, remain stable against this severe marginal misspecification, and their performance remains comparable to that under the correctly specified model (see Table~\ref{SIM2}).
\begin{table}[htbp]
\centering
\caption{ {Results for misspecified models I and II}}\label{SIM2}
\vspace{3mm}
\begin{tabular}{|l|cc|cc|}
\hline
 & \multicolumn{2}{c||}{\shortstack{Misspecified\\ case I}} & \multicolumn{2}{c|}{\shortstack{Misspecified\\ case II}} \\

\hline\hline
     & Mean & RMSE & Mean & RMSE  \\
\hline
    MLE   &  {0.3696} &  {0.1837} &  {0.3829} &  {0.1741} \\
  $\alpha$-MCDE    &  {0.4027} &  {0.1599} &  {0.4197} &  {0.1492} \\
 $\beta$-MCDE  &  {0.3988} &  {0.1629} &  {0.4146} &  {0.1520} \\
$\gamma$-MCDE &  {0.4030} &  {0.1643} &  {0.4307} &  {0.1494} \\
\hline
\end{tabular}
\end{table}


 {To evaluate the finite-sample performance of the proposed MCDE in higher dimensions, we conducted a simulation study for copulas linking $d=10,15,$ and $20$ variables. The results, summarized in Table \ref{tab:simulation_summary}, provide strong numerical validation for the practical utility of the MCDE framework, especially in scenarios common in modern data analysis. Under the correctly specified model, the semiparametric Maximum Likelihood Estimator (MLE) demonstrates its known asymptotic efficiency by consistently achieving a slightly lower Root Mean Squared Error (RMSE) than the $\beta$-MCDE, where we fix as $\beta=0.1$. For instance, at $d=20$, the MLE's RMSE is $0.0566$ compared to the $\beta$-MCDE's $0.0637$. This confirms the expected trade-off, where the robust estimator incurs a minor efficiency loss in an ideal setting. However, the crucial advantage of the robust approach becomes evident in the contaminated model. Here, the $\beta$-MCDE significantly outperforms the MLE across all dimensions. The MLE is heavily influenced by the contamination, exhibiting a large bias (e.g., $-0.3571$ for $d=20$). In contrast, the $\beta$-MCDE maintains a much smaller bias (e.g., $-0.1004$ for $d=20$). This superior robustness translates to a dramatically lower overall error. At $d=20$, the $\beta$-MCDE achieves an RMSE of $0.1155$, which is less than one-third of the MLE's RMSE of $0.3605$. Notably, this robustness gap not only persists but widens as the complexity of the problem increases. These findings confidently show that the proposed robust estimators offer a crucial advantage over traditional methods when dealing with imperfect data, and this advantage grows with the dimensionality of the problem.}

\begin{table}[htbp]
\centering
\caption{Simulation summary for MLE and $\beta$-MCDE performance under correctly specified and contaminated models. Results are based on 100 repetitions with a sample size of $N=2500$ and a true parameter value of $\theta=2.0$.}
\label{tab:simulation_summary}
\vspace{2mm}
\begin{tabular}{lllrrrr}
\hline
{Dimension} &  {Scenario} &  {Estimator} &  {Mean} &  {StdDev} &  {Bias} &  {RMSE} \\
\hline
$d=10$ & Correctly Specified  & MLE       & 1.9938 & 0.0530 & -0.0062 & 0.0533 \\
$d=10$ & Correctly Specified  & $\beta$-MCDE & 2.0008 & 0.0601 &  0.0008 & 0.0601 \\
$d=10$ & Contaminated         & MLE       & 1.6605 & 0.0536 & -0.3395 & 0.3437 \\
$d=10$ & Contaminated         & $\beta$-MCDE & 1.8565 & 0.0624 & -0.1435 & 0.1565 \\
\hline
$d=15$ & Correctly Specified  & MLE       & 1.9987 & 0.0500 & -0.0013 & 0.0500 \\
$d=15$ & Correctly Specified  & $\beta$-MCDE & 2.0134 & 0.0648 &  0.0134 & 0.0662 \\
$d=15$ & Contaminated         & MLE       & 1.6441 & 0.0528 & -0.3559 & 0.3598 \\
$d=15$ & Contaminated         & $\beta$-MCDE & 1.8746 & 0.0596 & -0.1254 & 0.1389 \\
\hline
$d=20$ & Correctly Specified  & MLE       & 1.9929 & 0.0561 & -0.0071 & 0.0566 \\
$d=20$ & Correctly Specified  & $\beta$-MCDE & 2.0064 & 0.0634 &  0.0064 & 0.0637 \\
$d=20$ & Contaminated         & MLE       & 1.6429 & 0.0496 & -0.3571 & 0.3605 \\
$d=20$ & Contaminated         & $\beta$-MCDE & 1.8996 & 0.0570 & -0.1004 & 0.1155 \\
\hline
\end{tabular}
\end{table}

\begin{figure}[!htbp]
\centering
\includegraphics[width=\textwidth]{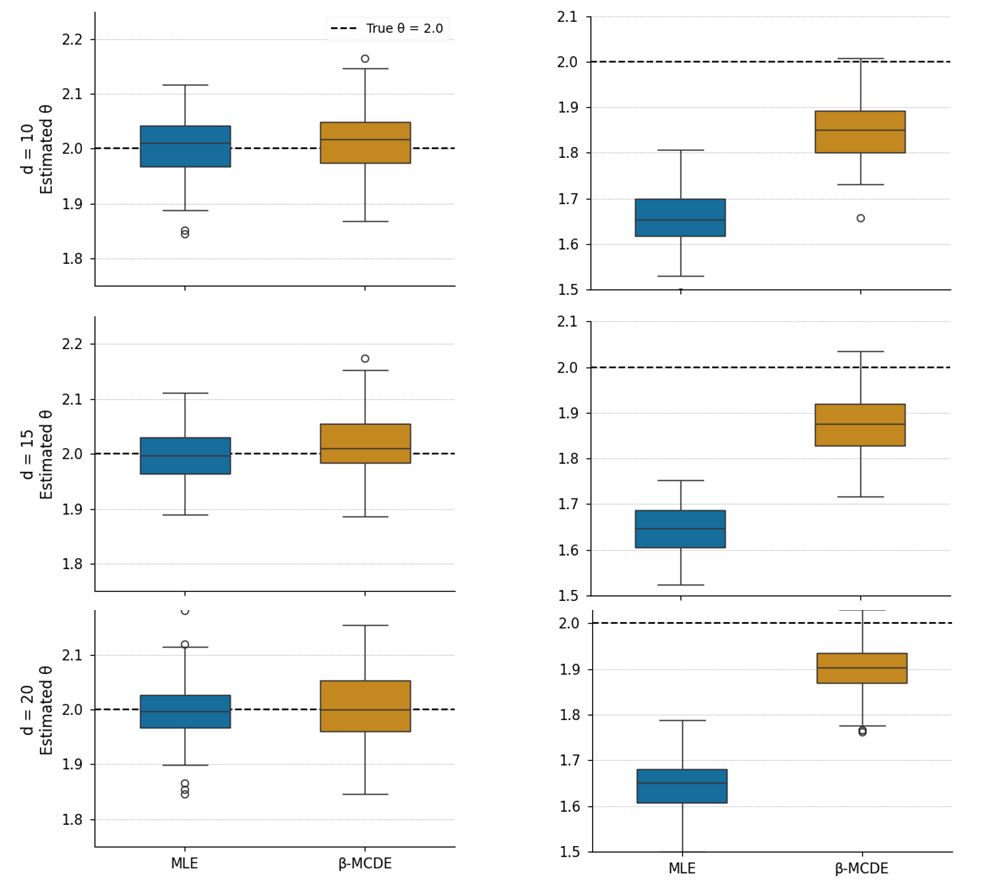}
\caption{Comparison of estimator in a higher-dimensional settings}
\label{SMI44}
\end{figure} 

{Next, we investigate the numerical performance of the data-driven selection for the exponent $\beta$ discussed in the preceding section. For this, we conduct a five-fold cross-validation to get cross-validated errors. Similarly, we consider the corrupted copula model with a heavy contamination of $\pi=0.1$ from a Student's t copula. Thus, a comparison among the MLE, the $\beta$-MCDE with fixed $\beta=1$ and $\beta=0.1$, and the optimal data-driven $\beta$-MCDE was conducted with 50 repetitions. The outputs are given in Table \ref{tab:sim_results_updated}.
As shown in Table \ref{tab:sim_results_updated}, the semiparametric MLE performs best in the correctly specified model in terms of RMSE (0.1315), which is expected due to its high efficiency. The robust MCDE methods exhibit a slightly higher RMSE, illustrating the classic trade-off between efficiency and robustness. In the misspecified model, however, the advantages of the robust data-driven approach become clear. The {Opt $\beta$-MCDE} achieves the lowest RMSE (0.1530), demonstrating a superior ability to handle the contamination. It provides a more accurate estimate than not only the heavily biased MLE (RMSE of 0.1761) but also the fixed-parameter estimators. Notably, while the $\beta$-MCDE with $\beta=0.1$ is also highly robust in this scenario, the Opt $\beta$-MCDE still performs slightly better, validating the cross-validation procedure. Conversely, the $\beta$-MCDE with $\beta=1$ is shown to be a poor choice for this type of contamination, performing even worse than the MLE. While even the robust methods are not perfectly accurate under heavy contamination, the key takeaway is their superior \textit{relative} performance. The MLE becomes significantly biased, whereas the data-driven optimal $\beta$-MCDE adapts to the contamination and degrades more gracefully, providing a more reliable estimate than both the MLE and the fixed-parameter MCDEs.}

\begin{table}[htbp]
\centering
\caption{K-fold Cross-Validation Results}
\label{tab:sim_results_updated}
\begin{tabular}{l cc cc}
\toprule
           & \multicolumn{2}{c}{\shortstack{Correctly specified\\ case}} & \multicolumn{2}{c}{\shortstack{Misspecified\\ case}} \\
           \cmidrule(r){2-3} \cmidrule(l){4-5}
Estimator  & Mean      & RMSE      & Mean       & RMSE      \\
\midrule
MLE                      & 0.5110 & 0.1315 & 0.3762 & 0.1761 \\
$\beta$-MCDE ($\beta=1$) & 0.5445 & 0.1537 & 0.3578 & 0.1898 \\
$\beta$-MCDE ($\beta=0.1$) & 0.5452 & 0.1382 & 0.3947 & 0.1556 \\
Opt $\beta$-MCDE       &  0.5462 & 0.1382 & 0.3982 & 0.1530 \\
\bottomrule
\end{tabular}
\end{table}

\begin{figure}[h]
\centering
\includegraphics[width=\textwidth]{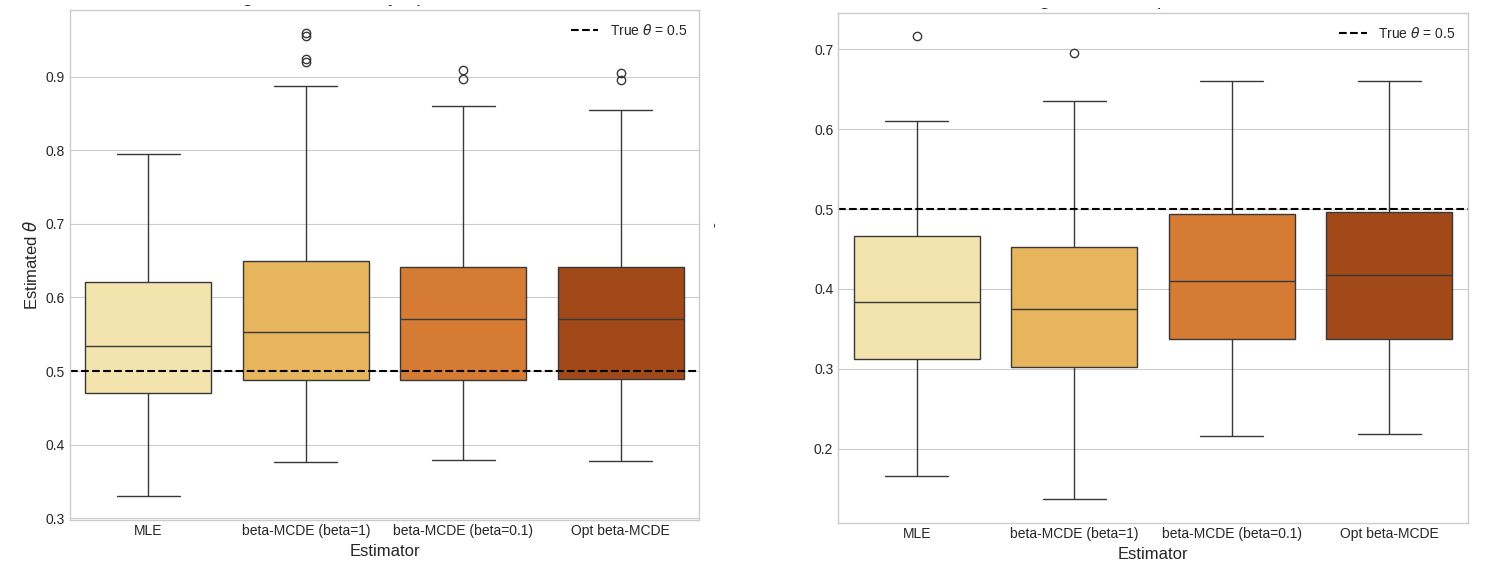}
\caption{Distribution of estimated $\theta$ under the correctly specified and misspecified  models}
\label{SMI33}
\end{figure}

\section{Concluding remarks}

This paper introduces and develops the concept of minimum copula divergence  as a robust estimation framework for analyzing dependence structures among variables. By isolating the copula component from marginal distributions, the proposed method effectively addresses challenges posed by non-linear or asymmetric relationships. This approach is not only theoretically sound but also highly versatile, as demonstrated through its potential applications in diverse fields such as finance, machine learning, and ecology. One of the primary contributions of this study is the formulation of a robust estimator that is resilient to outliers and model misspecifications. Theoretical results, coupled with numerical experiments, underscore the superiority of MCDE over conventional estimation methods under various contamination scenarios. The use of divergence-based metrics provides a flexible and robust alternative for modeling dependence structures, particularly in high-dimensional or noisy data environments. An additional advantage of the MCDE framework lies in its nonparametric nature, which leverages rank statistics to estimate dependence structures. By relying on rank-based measures, MCDE avoids assumptions about specific parametric forms of the marginal distributions, thereby enhancing its robustness and generalizability. This nonparametric property also makes MCDE particularly suitable for datasets where parametric assumptions are difficult to justify or validate. The implications of MCDE extend beyond its immediate statistical applications. Its ability to model intricate dependency patterns opens doors for innovations in areas where understanding complex relationships is critical. For example, in financial risk management, MCDE can improve portfolio optimization by accurately capturing tail dependencies. Similarly, in ecological studies, it can facilitate more reliable predictions of species distributions under uncertain conditions. MCDE's relationship with optimal transport theory presents another intriguing avenue for exploration. Optimal transport, a framework rooted in information geometry, provides a powerful tool for analyzing the geometry of probability distributions. See \cite{{amari-nagaoka2000},{eguchi1983},{eguchi1992geometry},{eguchi2006interpreting}} for extensive discussion and statistical applications. 
By aligning the objectives of MCDE with those of optimal transport, such as minimizing divergence in the distributional sense, 
future work could further enhance the theoretical foundation of robust dependency modeling. The coupling of MCDE with concepts from information geometry could yield deeper insights into the structure of dependence and improve its adaptability to complex, real-world data scenarios. Despite its strengths, the framework is not without limitations. 
 {As noted, the computational complexity of MCDE depends on the tractability of the copula CDF, which can be a challenge in high-dimensional settings for certain copula families like vine copulas.}
Moreover, the choice of divergence function and its impact on estimator performance require additional investigation. Addressing these issues will enhance the practical applicability of MCDE. Future research directions include exploring alternative divergence measures and extending the methodology to dynamic or time-varying copula models. Additionally, integrating MCDE with machine learning techniques could provide novel insights and improve predictive capabilities in data-intensive applications. In conclusion, this study establishes MCDE as a promising tool for robust estimation in dependency modeling. By bridging theoretical advancements and practical applications, it lays the groundwork for future innovations in robust statistical methods and their interdisciplinary applications.

\subsection*{Appendix 1}

\noindent
[1]. For the Clayton copula model, we express 
\[
\log C_\theta(u, v) = -{\theta} \log(u^{-1/\theta} + v^{-1/\theta}-1)
\]
and hence the partial derivative with respect to \(\theta\):
\[
\nabla_\theta \log C_\theta(u, v) = {\log(u^{-1/\theta} + v^{-1/\theta}-1)}
 - \frac{u^{-1/\theta} \log u + v^{-1/\theta} \log v}{\theta (u^{-1/\theta} + v^{-1/\theta}-1)}.
\]
We then consider the limit as \((u, v) \to (0, 0)\). By setting \(u = v\), we simplify the expression:
\[
\nabla_\theta \log C_\theta(u, u) = -{\log(2u^{-1/\theta}-1)}
 - \frac{2u^{-1/\theta} \log u}{2\theta (u^{-1/\theta}-1)}
\]
and
\[
C_\theta(u, u)^\alpha = (2 u^{-1/\theta}-1)^{-\alpha \theta}.
\]
Thus, the expression becomes:

\begin{align}\label{limitXI}
C_\theta(u, u)^\alpha \log C_\theta(u, u) =\Xi^{-\alpha\theta}\left\{-\log(\Xi)-\frac{1}{\theta^2}
\frac{\Xi+1}{\Xi}\log\Bigl(\frac{\Xi+1}{2}\Bigr)\right\}
\end{align}
where $\Xi=2u^{-1/\theta}-1$. As \(u \to 0\), \(\Xi \to \infty\) for $\alpha>0$, so the LHS in \eqref{limitXI} tends to 0.
 {This analysis along the path $u=v$ is done for analytical simplicity. The limiting behavior is consistent across other paths to the origin (e.g., $v=ku$ for $k>0$), thus the conclusion holds for any $(u, v)$ in a neighborhood of (0,0).}
Therefore, the limit is:
\[
{ \lim_{(u,v) \to (0,0)} |C_\theta(u,v)^\alpha \nabla_\theta \log C_\theta(u,v) |= \begin{cases} 0 & \text{if } \alpha > 0, \\ \infty & \text{if } \alpha = 0. \end{cases} }
\]

\bigskip

\noindent
[2]. For the Gumbel copula model, we express 
\[
\log C_\theta(u,v) = -\left( (-\log(u))^\theta + (-\log(v))^\theta \right)^{1/\theta}
\]
Hence,
\[
\nabla_\theta \log C_\theta(u,v) = a^{1/\theta} \left( \frac{\log a}{\theta^2} - \frac{a'}{\theta a} \right),
\]
where \( a = (-\log(u))^\theta + (-\log(v))^\theta \). By setting \( u = v \), we simplify the expression:
\[
\nabla_\theta \log C_\theta(u,u) = 2^{1/\theta} (-\log u) \left( \frac{\log 2 + \theta \log (-\log u)}{\theta^2} - \frac{\log (-\log u)}{\theta} \right)
\]
and
\[
C_\theta(u,u )^\alpha = \exp(-\alpha 2^{1/\theta} (-\log u))
\]
Therefore, the limit expression of $C_\theta(u,u)^\alpha \nabla_\theta \log C_\theta(u,u)$ is:
\[
\exp(-\alpha 2^{1/\theta} (-\log u)) \cdot 2^{1/\theta} (-\log u) \left( \frac{\log 2 + \theta \log (-\log u)}{\theta^2} - \frac{\log (-\log u)}{\theta} \right)
\]
as \( u \to 0 \), \( -\log u \to \infty \), and the exponential term \(\exp(-\alpha 2^{1/\theta} (-\log u))\) decays to 0 for \(\alpha > 0\), while the polynomial and logarithmic terms grow slower than the exponential decay.
For \(\alpha = 0\), the expression simplifies to:
\[
2^{1/\theta} (-\log u) \left( \frac{\log 2 + \theta \log (-\log u)}{\theta^2} - \frac{\log (-\log u)}{\theta} \right)
\]
which tends to infinity as \( u \to 0 \).  {As with the Clayton example, this result holds for other paths to the origin.}
Therefore, the limit is:
\[
{ \lim_{(u,v) \to (0,0)} C_\theta(u,v)^\alpha \nabla_\theta \log C_\theta(u,v) = \begin{cases} 0 & \text{if } \alpha > 0, \\ \infty & \text{if } \alpha = 0. \end{cases} }
\]

\bigskip

\noindent
[3]. For the Frank copula function:
\[ C_\theta(u,v) = -\frac{1}{\theta} \log\left[1 + \frac{(\exp(-\theta u) - 1)(\exp(-\theta v) - 1)}{\exp(-\theta) - 1}\right],
 \]
 we approximate \( C(u,v,\theta) \) for small \( u \) and \( v \):
\[ C_\theta(u,v) \approx \frac{\theta u v}{1 - \exp(-\theta)} \]
This implies
\[ \nabla_\theta \log C_\theta(u,v) \approx \frac{1}{\theta} - \frac{1}{\exp(\theta) - 1} \]
Now, we observe
\[ 
C_\theta(u,v)^\alpha \; \nabla_\theta \log C_\theta(u,v) \approx \left(\frac{\theta u v}{1 - \exp(-\theta)}\right)^\alpha \left( \frac{1}{\theta} - \frac{1}{\exp(\theta) - 1} \right) 
\]
for small \( u \) and \( v \). Therefore, 
\[
{\lim_{(u,v) \to (0,0)} C_\theta(u,v)^\alpha \; \nabla_\theta \log C_\theta(u,v) = \begin{cases} 0 & \text{if } \alpha > 0, \\ \frac{1}{\theta} - \frac{1}{\exp(\theta) - 1} & \text{if } \alpha = 0. \end{cases}} 
\]

\bigskip

\noindent
[4]. For the Joe copula function:
 \[ 
C_\theta(u, v) = 1 - \left[ (1 - u)^\theta + (1 - v)^\theta - (1 - u)^\theta (1 - v)^\theta \right]^{1/\theta} ,
\]
we approximate \( C_\theta(u,v,\theta) \) for small \( u \) and \( v \):
\[
C_\theta(u,v) \approx \theta u v
\]
Therefore,
\[
\nabla_\theta \log C_\theta(u, v) \approx \frac{1}{\theta}
\]
now, consider the product:
\[
C_\theta(u, v)^\alpha   \nabla_\theta \log C_\theta(u, v) \approx (\theta u v)^\alpha   \frac{1}{\theta} = \theta^{\alpha - 1} u^\alpha v^\alpha
\]
as \( (u, v) \to (0,0) \), \( u^\alpha v^\alpha \to 0 \) for \( \alpha > 0 \), and remains 1 for \( \alpha = 0 \).
In summary,
\[
{ \lim_{(u,v) \to (0,0)} C_\theta(u,v, )^\alpha \; \nabla_\theta \log C(u,v,\theta) = \begin{cases} 0 & \text{if } \alpha > 0, \\ \dfrac{1}{\theta} & \text{if } \alpha = 0. \end{cases} }
\]

\subsection*{Appendix 2}
 
The Gaussian copula can be written as
\(
C_{\theta}(u,v)  =  \Phi_{\theta}(x,y),
\)
where \(x =\Phi^{-1}(u),\,y =\Phi^{-1}(v)\). For large negative $x,y$, $\Phi_{\theta}(x,y)$ is very small, a standard Laplace-type asymptotic 
for the bivariate normal CDF shows that
\[
\Phi_{\theta}(x,y)
\approx
\frac{1}{2\pi\,\sqrt{1-\theta^2}}
 \exp \Bigl(-Q(x,y;\theta)\Bigr)
\]
up to polynomial factors in $|x|,|y|$, where 
\[
Q(x,y;\theta)
=
\frac{x^2 - 2\theta\,x\,y + y^2}{\,2\,(1-\theta^2)\,}.
\]
Then, for $(x,y)\to(-\infty,-\infty)$,
\(
\Phi_{\theta}(x,y) 
 \sim  \exp\bigl[-\,Q(x,y;\theta)\bigr].
\)
Thus, we obtain:
\[
\frac{\nabla_{\theta}\,\Phi_{\theta}(x,y)}{\Phi_{\theta}(x,y)}
 \sim 
-\,\nabla{\theta}Q(x,y;\theta),
\]
which is unbounded in absolute value as $x,y\to -\infty$. Therefore, 
\[
\frac{\nabla_{\theta}\,\Phi_{\theta}(x,y)}{\Phi_{\theta}(x,y)}
 \to \pm\infty
\quad\text{as }(x,y)\to(-\infty,-\infty),
\]
and hence
\[
\bigl|\nabla_{\theta} \log C_{\theta}(u,v)\bigr|
 = 
\frac{|\nabla_{\theta}\,\Phi_{\theta}(x,y)|}{\Phi_{\theta}(x,y)} 
 \to  \infty.
\]

\medskip
Noting \(C_{\theta}(u,v)^{\alpha}
\sim \exp\!\bigl[-\,\alpha\,Q(x,y;\theta)\bigr]
\) from the previous asymptotic,  
\[
C_{\theta}(u,v)^\alpha \bigl|\nabla_{\theta} \log C_{\theta}(u,v)\bigr|
\sim
\exp \bigl[-\,\alpha\,Q(x,y;\theta)\bigr]\times\Bigl| \nabla_{\theta}Q(x,y;\theta)
\Bigr|
\]
Therefore the product remains dominated by an exponentially decaying factor for any $\alpha>0$, 
while the derivative factor grows only polynomially. Hence 
\[
\lim_{(u,v)\to(0,0)}
C_{\theta}(u,v)^{\alpha}\,
\bigl|\nabla_{\theta}\log C_{\theta}(u,v)\bigr|
 =  0
\quad
\text{for all }\alpha>0.
\]

\subsection*{Appendix 3}
We prove Proposition \ref{prop1} as follows. By the definition of $\alpha$-MCDE $\hat{\bm\theta}_\alpha$,
\[
   \sum_{i=1}^n \bm S_\alpha(\bm U_i,\hat{\bm\theta}_\alpha)  ={\bf0}.
\]
The Taylor theorem yields
\[ 
 \sum_{i=1}^n \bigl\{\bm S_\alpha(\bm U_i,\bm\theta) +\nabla_{\bms\theta}\otimes\bm S_{\alpha}(\bm U_i,\bm\theta)\; (\hat{\bm\theta}_\alpha-\bm\theta)
\bigr\} =
o(\|\hat{\bm\theta}_\alpha-\bm\theta\|)
\]
Hence multiplying both sides by $1/\sqrt n$  yields
\[ 
\frac{1}{\sqrt{n}}\sum_{i=1}^n \bm S_{\alpha}(\bm U_i\bm\theta)+
\Bigl\{\frac{1}{n}\sum_{i=1}^n \nabla_{\bms\theta}\otimes\bm S_{\alpha}(\bm U_i,\bm\theta)\Bigr\}\; \sqrt{n}(\hat{\bm\theta}_\alpha-\bm\theta)=o_p(1)
\]
We observe that  {a first-order Taylor expansion gives}
\[ 
\bm S_{\alpha}(\bm u,\bm\theta)
 {=} \{C_{\bms\theta}(\bm u)-\hat C(\bm u)\}\;\nabla_{\bms\theta}\log C_{\bms\theta}(\bm u)  {+ O_p(n^{-1})}
\]
and hence 
\[ 
\frac{1}{\sqrt{n}}\sum_{i=1}^n\bm S_{\alpha}(\bm U_i\bm\theta)
 {=} \sqrt{n}\int \{C_{\bms\theta}(u)-\hat C(\bm u)\}\;\nabla_{\bms\theta}\log C_{\bms\theta}(u){\rm d}C_{\bms\theta}(\bm u)+o_p(1).
\]
By the continuous mapping theorem,
\[ 
\sqrt{n}\int \{C_{\bms\theta}(u)-\hat C(\bm u)\}\;\nabla_{\bms\theta}\log C_{\bms\theta}(u){\rm d}C_{\bms\theta}(\bm u)    \Longrightarrow   
 \int  W(\bm{u})\;\nabla_{\bms\theta}\log C_{\bms\theta}(u){\rm d}C_{\bms\theta}(\bm u),
\]
which is equal to a normal distribution with mean $\bf0$ and covariance $B_0(\bm\theta)$, where
$B_0(\bm\theta)$ is defined in \eqref{B-theta} with $x=0$. This is because the empirical copula converges weakly to a mean-zero Gaussian process:
\[
  \sqrt{n}\,\bigl(\widehat{C}(\bm{u}) - C(\bm{u})\bigr)
    \Longrightarrow   
  W(\bm{u}),
  \quad \bm{u}\in [0,1]^d,
\]
where \(W\) is  a general Gaussian process with the mean \(\bf0\) and covariance given by  
\[
  \mathbb{E}\bigl[\,W(\bm{u})\,W(\bm{v})\bigr]
   = 
  \Gamma\bigl(\bm{u},\,\bm{v}\bigr)
   = 
  C\!\bigl(\bm{u} \wedge \bm{v}\bigr)
   - 
  C(\bm{u})\,C(\bm{v}),
\]
where \(\bm{u} \wedge \bm{v}=\bigl(\min(u_1,v_1),\dots,\min(u_d,v_d)\bigr)\). We observe, as $n\to\infty$
\[
\frac{1}{n}\sum_{i=1}^n \nabla_{\bms\theta}\otimes\bm S_{\alpha}(\bm U_i,\bm\theta)
\longrightarrow
\int C_{\bms\theta}(\bm u)
\nabla_{\bms\theta}\log C_{\bms\theta}(\bm u)
\otimes\nabla_{\bms\theta}\log C_{\bms\theta}(\bm u){\rm d}C_{\bms\theta}(\bm u)
\]
in probability, which is equal to $A_0(\theta)$ in \eqref{A-theta}. In accordance with these,
$\sqrt{n}(\hat{\bm\theta}_\alpha-\bm\theta)  \Longrightarrow N({\bf0},\Sigma_0)$ noting
\[
\sqrt{n}(\hat{\bm\theta}_\alpha-\bm\theta)=
\sqrt{n}A_0(\bm\theta)^{-1}\int \{C_{\bms\theta}(\bm u)-\hat C(\bm u)\}\;\nabla_{\bms\theta}\log C_{\bms\theta}(\bm u){\rm d}C_{\bms\theta}(\bm u)+o(1) .
\]
An argument for the $\beta$-MCDE similar to that for the $\alpha$-MCDE yields
 \[ 
\sqrt{n}(\hat{\bm\theta}_\beta-\bm\theta)
 \Longrightarrow   A_\beta(\bm\theta) ^{-1}  
 \int  W(\bm{u})C_{\bms\theta}(\bm u)^\beta\;\nabla_{\bms\theta}\log C_{\bms\theta}(\bm u){\rm d}C_{\bms\theta}(\bm u)
\]
noting
\[
\frac{1}{n}\sum_{i=1}^n \nabla_{\bms\theta}\otimes\nabla_{\bms\theta}\bm S_{\beta}(\bm U_i,\bm\theta)
\longrightarrow
A_\beta(\theta)
\]
in probability. Hence, $\sqrt{n}(\hat{\bm\theta}_\beta-\bm\theta)  \Longrightarrow N({\bf0},\Sigma_\beta)$.
Similarly, for the $\gamma$-MCDE,
 \[ 
\sqrt{n}(\hat{\bm\theta}_\gamma-\bm\theta)
 \Longrightarrow   A_\gamma(\bm\theta) ^{-1}  
 \int  W(\bm{u})C_{\bms\theta}(\bm u)^\gamma\;\nabla_{\bms\theta}\log C_{\bms\theta}(\bm u){\rm d}C_{\bms\theta}(\bm u)
\]
since $\hat w_{\gamma\bms{\theta}}$ coverges to $1$ in probability. This concludes $\sqrt{n}(\hat{\bm\theta}_\gamma-\bm\theta)  \Longrightarrow N({\bf0},\Sigma_\gamma)$.
The proof is complete.

\subsection*{Appendix 4}
We prove Proposition \ref{prop2}. \noindent
By the definition \eqref{alpha-est},
\[
\|\bm S_\alpha(\bm u,\bm{\theta})\|= \frac{1}{|1-\alpha|}
 C_{\bms\theta}(\bm u)^\alpha \bigl| 
C_{\bms\theta}(\bm u)^{1-\alpha}-{\hat C}(\bm u)^{1-\alpha}
\bigr| \cdot \bigl\| \nabla_{\bms\theta} \log C_{\bms\theta}(\bm u) \bigr\|.
\]
If $0<\alpha<1$, then
\[
\|\bm S_\alpha(\bm u,\bm{\theta})\|\leq \frac{1}{1-\alpha}
 C_{\bms\theta}(\bm u)^\alpha  \big\| \nabla_{\bms\theta} \log C_{\bms\theta}(\bm u) \big\|.
\]
due to $\bigl| C_{\bms\theta}(\bm u)^{1-\alpha}-{\hat C}(\bm u)^{1-\alpha}\bigr|\leq1$.
This implies $\bm S_\alpha(\bm u,\bm{\theta})$ is uniformly bounded in $[0,1]^d$ from the assumption of the power-boundedness for $C_{\bms\theta}(\bm u)$. Similarly, for $\beta>0$, 
\[
\|\bm S_\beta(\bm u,\bm{\theta})\|\leq 
 C_{\bms\theta}(\bm u)^\beta  \big\| \nabla_{\bms\theta} \log C_{\bms\theta}(\bm u) \big\|.
\]
This implies the uniform boundedness for $\bm S_\beta(\bm u,\bm{\theta})$  from the assumption. Finally, for $\gamma>0$, 
\[
\|\bm S_\gamma(\bm u,\bm{\theta})\|\leq 
 C_{\bms\theta}(\bm u)^\gamma  \big\| \nabla_{\bms\theta} \log C_{\bms\theta}(\bm u) \big\|.
\]
since $\big| \hat w_{\gamma,{\bms\theta}}\; {C_{\bms\theta}(\bm u)} -{\hat{C}(\bm u)}\big|\leq 1$.
This implies the uniform boundedness for $\bm S_\gamma(\bm u,\bm{\theta})$  from the assumption.


\begin{thebibliography}{99}
	\bibitem[Alquier et al.(2023)]{alquier2023}
Alquier, P., Ch\'{e}rief-Abdellatif, B. E., Derumigny, A., and Fermanian, J. D. (2023).
\newblock Estimation of copulas via maximum mean discrepancy. 
\newblock {\em Journal of the American Statistical Association}, 118(543), 1997-2012.
	\bibitem[Amari and Nagaoka(2000)]{amari-nagaoka2000}
	Amari, S. I., and  Nagaoka, H.  
	\newblock {\em Methods of information geometry (Vol. 191)}.
\newblock American Mathematical Soc., 2000.
	
	\bibitem[Basu et al. et al.(1998)]{basu1998robust}
	Basu, A., Harris, I. R.,  Hjort, N. L. and Jones, M. C.
	\newblock Robust and efficient estimation by minimising a density power divergence.
\newblock {\em Biometrika}, 85(3):549--559, 1998.
	
	\bibitem[Bevilacqua et al.(2024)]{bevilacqua2024}
	Bevilacqua, M., Alvarado, E., and Caamaño-Carrillo, C. (2024).
\newblock  A flexible Clayton-like spatial copula with application to bounded support data.
\newblock {\em Journal of Multivariate Analysis}, 201, 105277.

	\bibitem[Charpentier and Segers(2009)]{charpentier(2009)}
Charpentier, A., and Segers, J. (2009).
\newblock Tails of multivariate Archimedean copulas. 
\newblock {\em Journal of Multivariate Analysis}, 100(7), 1521-1537.
	\bibitem[Chen et al.(2006)]{chen2006efficient}
     {Chen, X., Fan, Y., and Tsyrennikov, V. (2006).
\newblock Efficient estimation of semiparametric multivariate copula models. 
    \newblock {\em Journal of the American Statistical Association}, 101(475), 1228-1240.}
		
	\bibitem[Durante and Sempi(2016)]{durante2016}
Durante, F., and Sempi, C. (2016).
Principles of copula theory (Vol. 474). 
Boca Raton, FL: CRC press.


	\bibitem[Eguchi(1983)]{eguchi1983}
	Eguchi, S. 
\newblock Second order efficiency of minimum contrast estimators in a curved exponential family.
\newblock {\em The Annals of Statistics}, 793-803, 1983.
	
	\bibitem[Eguchi(1992)]{eguchi1992geometry}
	Eguchi, S.
	\newblock Geometry of minimum contrast.
\newblock {\em Hiroshima Mathematical Journal}, 22(3):631--647, 1992.
	
	\bibitem[Eguchi and Copas(2006)]{eguchi2006interpreting}
	Eguchi, S. and Copas, J.
	\newblock Interpreting Kullback--Leibler divergence with the Neyman--Pearson
	lemma.
\newblock {\em Journal of Multivariate Analysis}, 97(9):2034--2040, 2006.
	
	
	\bibitem[Eguchi(2024)]{eguchi2024}
	Eguchi, S. 
	\newblock Minimum Gamma Divergence for Regression and Classification Problems.
\newblock {\em arXiv preprint} arXiv:2408.01893, 2024.
	
	
	\bibitem[Eguchi and Komori(2022)]{eguchi2022minimum}
	Eguchi, S. and Komori, O.
	\newblock {\em Minimum divergence methods in statistical machine learning: From an Information Geometric Viewpoint}.
\newblock {\em Springer Japan KK}, 2022.
	
	
	\bibitem[Eguchi et al.(2011)]{eguchi2011projective}
	Eguchi, E., Komori, O. and Kato, S.
	\newblock Projective power entropy and maximum tsallis entropy distributions.
\newblock {\em Entropy}, 13(10):1746--1764, 2011.

	
	\bibitem[Fujisawa and Eguchi(2008)]{Fujisawa2008}
	Fujisawa, H. and Eguchi, S.
	\newblock Robust parameter estimation with a small bias against heavy
	contamination.
\newblock {\em Journal of Multivariate Analysis}, 99:2053--2081, 2008.

\bibitem[Genest et al.(1995)]{Genest1995}
 {Genest, C., Ghoudi, K., and Rivest, L. P. (1995).
\newblock A semiparametric estimation procedure of dependence parameters in multivariate families of distributions.
\newblock {\em Biometrika}, 82(3), 543-552.}
	
\bibitem[Genest and Rivest(1993)]{Genest1993}
Genest, C., and Rivest, L. P. (1993).
\newblock Statistical inference procedures for bivariate Archimedean copulas. 
\newblock {\em Journal of the American statistical Association}, 88(423), 1034-1043.
\bibitem[Ghoudi and Rémillard(2004)]{ghoudi2004empirical}
 {Ghoudi, K., and Rémillard, B. (2004). 
\newblock Empirical processes based on pseudo-observations II: the multivariate case.
\newblock In {\em Asymptotic methods in stochastics} (pp. 381-406). American Mathematical Society.}


\bibitem[Google Finance(2024)]{google-finance2024}
Google Finance.
2024. 
\newblock Historical Price Data for Microsoft (MSFT) and Apple (AAPL)." Accessed January 20, 2025. 
\newblock https://www.google.com/finance.

\bibitem[Hofert et al.(2019)]{hofert2019elements}
 {Hofert, M., Kojadinovic, I., Mächler, M., and Yan, J. (2019). 
\newblock {\em Elements of copula modeling with R}. 
\newblock Springer.}


	\bibitem[Joe(2014)]{joe2014} 
Joe, H. (2014). 
\newblock  Dependence modeling with copulas. 
\newblock   CRC press.

	\bibitem[Jones et al.(2001)]{jones2001}
   Jones, M. C., Hjort, N. L., Harris, I. R., and Basu, A. (2001). 
\newblock  A comparison of related density-based minimum divergence estimators. 
\newblock  {\em Biometrika}, 88(3), 865-873.

\bibitem[Kato et al.(2022)]{kato2022}
Kato, S., Yoshiba, T., and Eguchi, S. (2022). 
\newblock  Copula-based measures of asymmetry between the lower and upper tail probabilities. 
\newblock  {\em Statistical Papers}, 63(6), 1907-1929.


\bibitem[De Keyser and Gijbels(2024)]{de-keyser2024}
De Keyser, S., and Gijbels, I. (2024).
\newblock  Parametric dependence between random vectors via copula-based divergence measures.
\newblock  {\em Journal of Multivariate Analysis}, 105336.

\bibitem[McNeil et al.(2015)]{mcneil2015}
 {McNeil, A. J., Frey, R., and Embrechts, P. (2015).
\newblock {\em Quantitative risk management: Concepts, techniques and tools}. 
\newblock Princeton university press.}

\bibitem[Nelsen(2006)]{nelsen2006}
Nelsen, R. B. (2006).
\newblock {\em An introduction to copulas.} 
\newblock Springer.

\bibitem[Tsukahara(2005)]{tsukahara2005}
 {Tsukahara, H. (2005). 
\newblock Semiparametric estimation in copula models.
\newblock {\em The Canadian Journal of Statistics}, 33(3), 357-375.}
	
\end{thebibliography}
\end{document}